\documentclass[usenatbib]{mn2e} 
\usepackage{epsfig}
\usepackage{amsmath}
\def\gsim{\mathrel{\raise0.35ex\hbox{$\scriptstyle >$}\kern-0.6em
\lower0.40ex\hbox{{$\scriptstyle \sim$}}}}
\def\lsim{\mathrel{\raise0.35ex\hbox{$\scriptstyle <$}\kern-0.6em
\lower0.40ex\hbox{{$\scriptstyle \sim$}}}}

\def\som{\text{M}_{\odot}}

\title[CLUE]{CFHT Legacy Ultraviolet Extension (CLUE): Witnessing Galaxy Transformations up to 7 Mpc from Rich Cluster Cores}

\author[Lu et al.]{Ting Lu\thanks{E-mail: tinglu@phys.ethz.ch}$^{1,2}$, David G. Gilbank$^{1,3}$, Sean L. McGee$^{1,4}$, Michael L. Balogh$^{1}$, 
\newauthor
Sarah Gallagher$^{5}$\\
\\
$^{1}$Department of Physics and Astronomy, University of Waterloo, Waterloo, Ontario, N2L 3G1, Canada\\
$^{2}$Institute of Astronomy, ETH Zurich, CH-8093 Zurich, Switzerland\\
$^{3}$South African Astronomical Observatory, Observatory Road, Observatory, 7925, South Africa\\
$^{4}$Department of Physics, Durham University, South Road, Durham, DH1 3LE, UK\\
$^{5}$Department of Physics and Astronomy, University of Western Ontario, London, Ontario, N6A 3K7, Canada
}

\date{\today}

\begin{document}

\maketitle

\begin{abstract}
Using the optical data from the Wide component of the CFHT Legacy
Survey, and new ultraviolet data from GALEX, we study the
colours and specific star formation rates (SSFR) of $\sim 100$ galaxy
clusters at $0.16<z<0.36$, over areas extending out to radii of $r\sim 7$
Mpc.  We use a multicolour, statistical
background subtraction method to study the galaxy population at this
radius; thus our results pertain to those galaxies which constitute an
excess over the average field density.
We find that the
average SSFR, and its distribution, of the star-forming galaxies (with
SFR$>0.7\ \som $/yr at $z\sim 0.2$ and  SFR$>1.2\ \som $/yr at $z\sim 0.3$)
have no measurable 
dependence on the cluster-centric radius, and are
consistent with the field values.  However, the fraction of galaxies
with SFR above these thresholds, and the fraction of optically
blue galaxies, 
are lower for the overdense galaxy population in the cluster outskirts compared with the average field value, at all
stellar masses $\text{M}_* >10^{9.8}\ \som $ and at all radii out to at least $7$
Mpc.  Most interestingly, the fraction of blue
galaxies that are forming stars at a rate below our UV detection limit is
much higher in all radial bins around our cluster sample, compared with
the general field value.  This is most noticeable for massive galaxies
$\text{M}_* >10^{10.7}\ \som$; while almost all blue field galaxies of this mass
have detectable star formation, this is true for less than 20\% of the blue cluster
galaxies, even at 7 Mpc from the cluster centre. Our results support a
scenario where galaxies are pre-processed in locally overdense regions,
in a way that reduces their SFR  below our UV detection limit, but not to zero. 

\end{abstract}

\begin{keywords}
galaxies: clusters: general, galaxies: evolution
\end{keywords}

\section{Introduction}

A lot has been learned about the star formation history of our Universe since early studies showed that the global star formation rate has declined by about a factor of 10 since $z\sim 1$ \citep[e.g][]{lilly96,madau96}. With decreasing redshift, the star formation rate of  star-forming galaxies has been decreasing \citep[e.g.][]{bell05,noeske07}, with an increase in the number of passive galaxies  \citep{pozzetti09}. The study by \cite{bell07} showed that a transformation from blue galaxies to red ones is needed, so that the stellar mass of today's blue galaxies is not overproduced.

The origin of this decline of the star formation rate and transformation remains unclear. One possible interpretation is that it is linked to the changing environment of galaxies.  It is known that in dense regions such as the cores of galaxy clusters, the population is dominated by galaxies with red colours and low average star formation rates \citep{balogh04b,Weinmann2006,Haines2006,kimm09}. Since, under the hierarchical paradigm, galaxy clusters grow by accreting galaxies that are generally star-forming, a transformation must have happened to quench the star formation. However, how and where this happens remains elusive. Studies have shown that the suppression of star formation is not restricted to cluster cores.  For example, \cite{G'omez2003}, \cite{balogh97,balogh98} and \cite{lewis02} detected a lower fraction of star-forming galaxies relative to the field beyond $\sim 2$ virial radii. 

Despite the low star-forming fraction in clusters, the detection of the change of star formation rate within the star-forming population itself is still an unsettled issue. Some studies found that the specific star formation rate of star-forming galaxies is the same  in the lowest and highest density regions  \citep{peng10},  in groups and in the field \citep{mcgalex10}, and in different regions in a $z\sim 0.2$ supercluster \citep{biv11}; but (for example) the study by \cite{vul2010} found a lower SFR of star-forming galaxies in $z\sim 0.5$ clusters than in the field.

An interesting place to look for the transformations is the outskirt regions of clusters. Simulations have shown that clusters are surrounded by large-scale structures such as filaments and sheets, and galaxies are accreted mainly along these structures \citep{bond96,colberg99}. These accretion zones are where galaxies reside  before they reach the cluster cores, and thus might be an attractive candidate for where the transformation of galaxies happens. However, studying the infall region of clusters is very difficult due to the low density contrast with the foreground/background field, and thus the exploration has only begun relatively recently. Studies focusing on two intermediate redshift clusters and the Shapley supercluster found evidence of obscured star formation and the transformation of spiral galaxies into S0 in infalling groups \citep[e.g.][]{geach06,Moran2007,haines11}.

In this paper, we use data from our CFHT Legacy Ultraviolet Extension (CLUE), to study the star formation properties of a large sample of clusters,  from the cluster core out to $\sim 7$ Mpc (the typical virial radius of clusters in our sample is $\sim 1-2$ Mpc).  We examine these properties as a function of stellar mass, as it has become clear that stellar mass is one of the key parameters that determines the properties of galaxies \citep{kau03}. In Section \ref{alldat}, we describe the data sets, and the cross-matching of the optical and UV catalogues. We describe our cluster and field samples, and the background subtraction procedure in Section \ref{samp}. The stellar mass and SFR estimates are described in Section \ref{anal}. We present our results on the star formation rate and fraction of blue/red galaxies as a function of cluster-centric radius  in Section \ref{gres}. We discuss the implications of our results in Section \ref{gdis}, and conclude in Section  \ref{con}.

We assume a cosmology 
with $\Omega_m=0.3$, $\Omega_{\Lambda}=0.7$ and
$H_o$=70 km s$^{-1}$ Mpc$^{-1}$ throughout. All magnitudes are in the AB system unless otherwise specified.

\section{Data}\label{alldat}
The data  we used in this study are from the Wide component of the CFHTLS  optical survey and our extended GALEX coverage over the CFHTLS fields. We describe them separately below.

\subsection{Optical Data}\label{lsdat}
The cluster sample we used in this study is detected from the  CFHTLS Wide survey, which is  a joint Canadian and French imaging survey in  $u^*$, $g'$, $r'$, $i'$ and $z'$ filters using the wide-field imager MegaCam. The survey is now complete,  covering a total of 171 square degrees, composed of single pointings each with a field of view of 1x1 square degree. The total exposure time are 6000s in $u^*$, 2500s in $g'$, 2000s in $r'$, 4300s in $i'$, and 7200s in $z'$. For extended sources, the 100 per cent completeness limit is $\sim 25$ mag in $u^*$, $\sim 25$ mag in $g'$, $\sim 24$ mag in $r'$, $\sim 24$ mag  in $i'$, and $\sim 23$ mag  in $z'$.   More details of the data sets are provided in \cite{lu09}, and thus here we only point out a few improvements.

One of the improvements is that  we used the stellar locus of stars
from SDSS to calibrate the colours of each individual CFHTLS
pointing. The $(g'-r')$ and $(r'-z')$ colours of stars in each pointing
are forced to agree with a reference set that is calibrated against SDSS stars but remains on the native MegaCam system (see \citealt{cali} for details).  This reduces the scatter in colour  when we stack a large number of clusters together. To calibrate the magnitude in each filter, we hold the $r'-$band magnitude fixed, and adjust the zeropoint of the other filters, on a field-by-field basis, to match the colour distribution of the stellar calibration set. This is because  the $r'-$band magnitude from Terapix, calibrated internally with respect to some reference pointings through overlap regions, has small variation between pointings, and is more accurate than the calibration against 2MASS, as described in \cite{cali}. The dispersion of the final calibrated magnitude among all pointings is  $\sim 0.01$ mag, similar in all filters. The other improvement is that we excluded regions around bright stars, which would lead to photometry with larger uncertainty and an underestimate of  the galaxy number density in these regions. This second effect has the most impact when we study the properties of galaxies far from cluster centres, where the density is low.

\subsection{GALEX Data}\label{gdat}

\begin{figure} 
\includegraphics[width=0.5\textwidth]{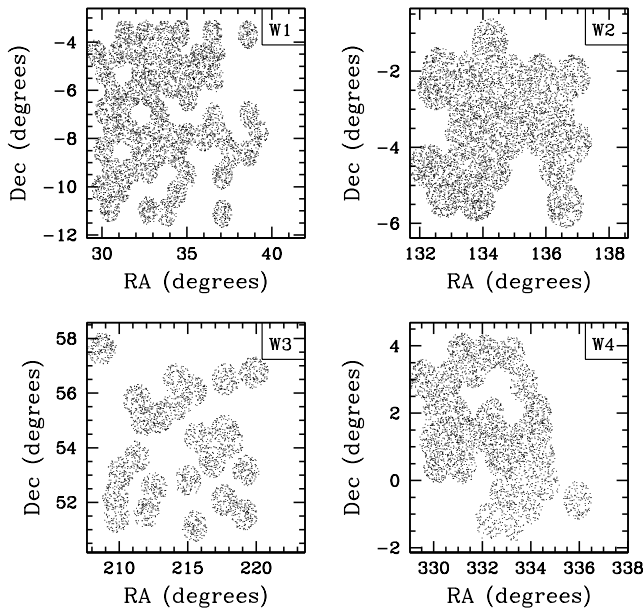} \caption[Current GALEX NUV coverage over the four CFHTLS Wide fields]{The current GALEX NUV coverage over the four CFHTLS Wide fields, including both our GI data and archival data. Points show objects brighter than NUV=23.0 mag. \label{cov}}
\end{figure}

In Cycle 5 of the GALEX Guest Investigator Programme, we proposed extended GALEX NUV coverage over the whole CFHTLS Wide fields  (GI5-28)\footnote{The full catalogues  will be published online once the survey is complete; in the meantime the current catalogues are available upon request.}, to the same depth as the Medium Imaging Survey ($\sim$1500s). The data collection is only partially complete, currently covering approximately 80 square degrees of the sky (about 50 per cent of the legacy field). In addition, we include existing archival data (57 pointings) over the CFHTLS Wide fields with exposure time greater than 1500s, which brings the total area with GALEX coverage to 110 square degrees.  For  multiple observations taken at the same position, the one with deeper exposure is used.  Figure \ref{cov} shows all the data we have in hand that are used in this work.

\begin{figure} 
\includegraphics[width=0.45\textwidth]{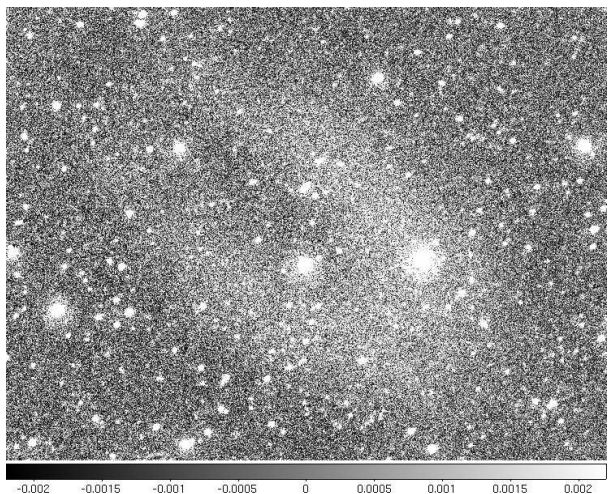} \caption[An example of the regions with residual elevated background in GALEX NUV images]{An example of the regions with residual elevated background in GALEX NUV images, which causes problems for both object detection and photometry measurement. The size of this region is about 0.15 x 0.15 square degree. The unit on the colourbar is counts per second per pixel. \label{Ushape}}
\end{figure}

We make use of the photometry provided in the NUV catalogue  measured using SExtractor \citep{sextractor}
 by the standard GALEX pipeline \citep{Morrissey2007}. The standard pipeline uses  a sophisticated method to correctly estimate the background in low-count regions. However, in some regions, there is residual elevated background,  likely due to reflection from bright stars. An example is shown in Figure \ref{Ushape}.  These elevated backgrounds cause  problems in both object detection and photometry measurements. Therefore, we mask out these regions when performing the analysis. The total area masked out is about 0.3 square degrees, which is insignificant compared to the whole coverage; however it can become more important when considering individual clusters.    

The full width half maximum (FWHM) of the GALEX point spread function (PSF)  is $\sim 4-5$ arcsec, and the field of view is circular, with a radius of  $\sim$0.6 degree. To avoid the slightly degraded astrometry and photometry near the edge of each pointing, we only include objects within a radius of 0.58 degree from the field centre. As  there are duplicate objects in the overlaps among adjacent pointings, we keep the one that is closest to the centre of the pointing it comes from.

\begin{figure} 
\includegraphics[width=0.5\textwidth]{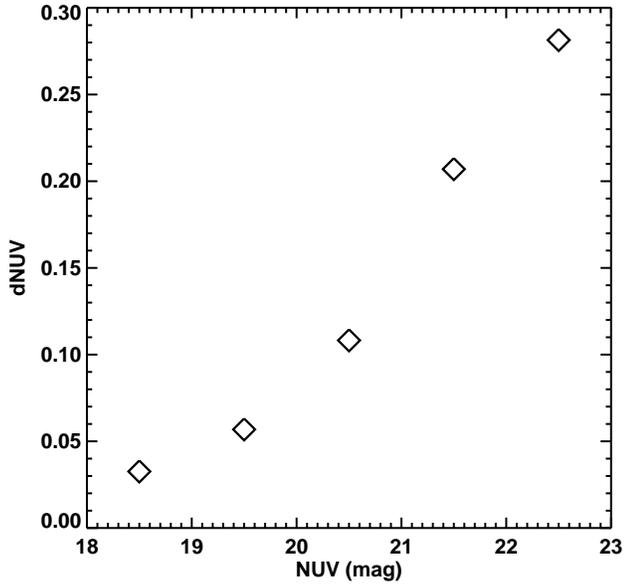} \caption{The standard deviation of the magnitude differences of duplicate objects measured from overlapping regions in GALEX NUV images, as a function of NUV magnitude.\label{err}}
\end{figure}

\begin{figure} 
\includegraphics[width=0.5\textwidth]{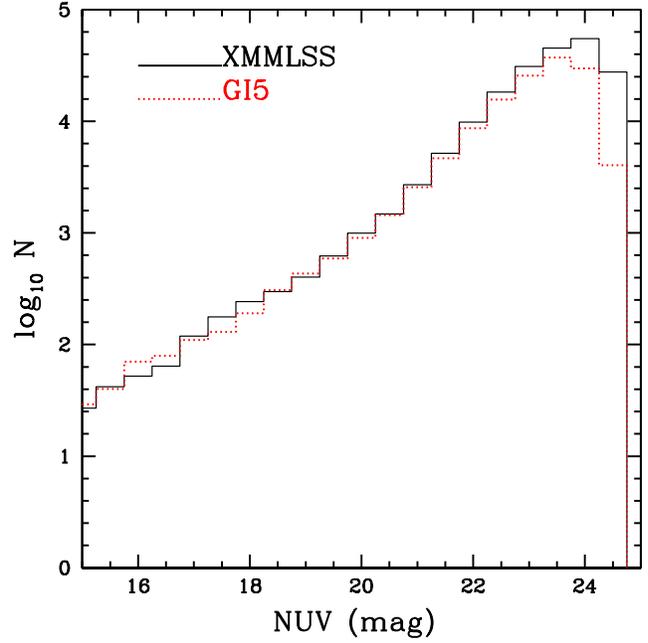}
\includegraphics[width=0.5\textwidth]{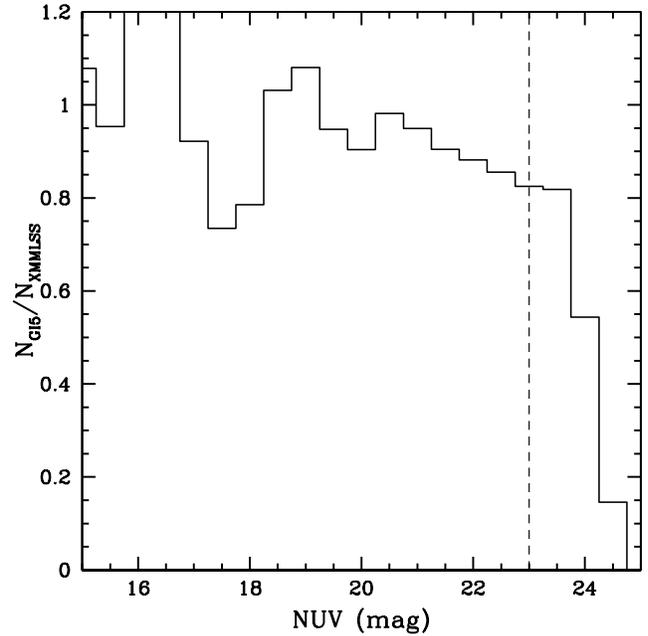} \caption[Number counts and completeness limit of our GI5 data]{Top panel: number counts as a function of magnitude (normalized to per tile) from our GI5 data (red dotted histogram) and the deeper data in the XMMLSS fields (black solid line). Bottom panel: completeness of our GI5 data as a function of magnitude. In this work, we limit our sample to those brighter than NUV=23.0 mag, which is the $\sim 80$ per cent completeness limit. \label{g_nct}}
\end{figure}

We use {\sc mag\_auto} as the total magnitude. We correct for Galactic
extinction using the relation between extinction and the reddening
determined by \cite{Wyder2007}, where the reddening E(B-V) is
calculated using the \cite{ext} dust map. Since the PSF of GALEX data
is much broader that of the CFHTLS data, when measuring  NUV-optical colours  we use {\sc mag\_auto} in both catalogues, instead of aperture magnitudes as for the optical-optical colour measurements. We estimate the NUV magnitude uncertainty using duplicate detections from overlapping pointings (note this gives an upper limit of the uncertainty because the overlapping regions are on the edge of each pointing, where the photometry is most uncertain). The standard deviation of the magnitude difference between duplicate objects is plotted in Figure \ref{err} as a function of magnitude. To estimate the completeness of our GI5 data, we compare the number counts from our GI5 data with that from the Deep Imaging Survey (DIS) in the XMM-Newton Large-Scale Strcture (XMMLSS) Survey fields. In the top panel of Figure \ref{g_nct}, the red dotted histogram shows the number count as a function of magnitude (normalized to per tile) from our GI5 data, and the black solid line shows the number count from the deeper data in the XMMLSS fields. In the bottom panel, we plot the ratio of the number counts from GI5 data to that from the XMMLSS fields, which shows that the $\sim 80$ per cent completeness of our GI5 data is about NUV=23.0 mag. Therefore, for our analysis here, we only consider objects brighter than this limit.

\subsection{Cross-matching Catalogues}\label{cross}

Since the CFHTLS data are much deeper than our GALEX data, essentially all NUV detected sources should have a match in the optical catalogue, except for extremely blue, optically faint galaxies ($NUV-u^*\lsim -3$ and $u^*\gsim 26$) that are not of interest here. Therefore, for each NUV source, we search the optical catalogue (without any magnitude cut) for matches within a radius of 4 arcsec (the optimal matching radius as discussed in \citealt{budavari09}). In $\sim 40$ per cent of cases, there are multiple candidate optical matches for one NUV source. 
To deal with this, we take the closest match, unless there is a second
candidate within 1 arcsec of it. This occurs for about 35 per cent of
the cases with multiple candidate matches (so 14 per cent of all galaxies). In such cases, we use the colours to help identify the most likely counterpart, under the assumption that the most likely match is the one that has the most common colour for a galaxy of its magnitude. This procedure is described in the Appendix.

We restrict the final matched catalogue to objects that are optically flagged as galaxies that are in the non-masked region (both around bright stars and regions with elevated background in the GALEX data,  see Sections \ref{lsdat} and \ref{gdat}), with SExtractor flag$<=$3 (i.e. excluding objects close to image edge, with corrupted aperture or with at least one pixel saturated) in all filters ($u^*,g',r',i',z'$ and $NUV$).

\section{Samples and Background Subtraction}\label{samp}
\subsection{Cluster Sample}
\begin{figure}
\includegraphics[width=0.5\textwidth]{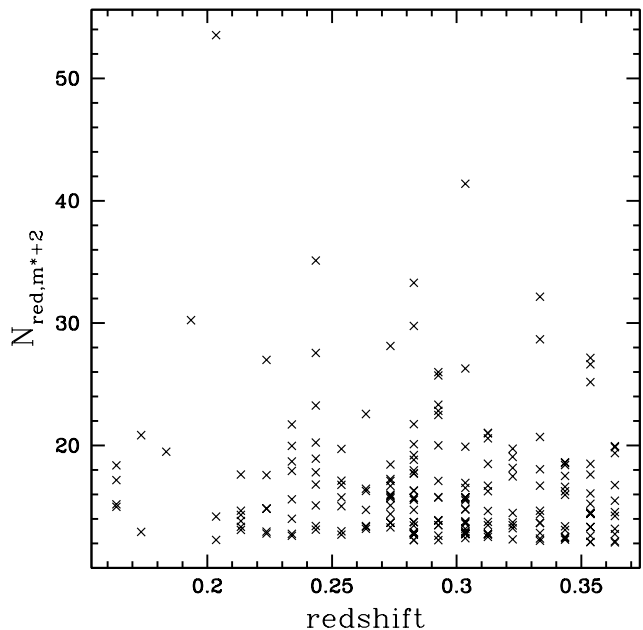}\caption{Richness vs. redshift of the clusters in our sample,  where the richness  $N_{red,m^*+2}$ is defined as the number of red-sequence galaxies brighter  than $m^*+2$, within a radius of 0.5 Mpc from the cluster centre.}\label{richdis}
\end{figure}

Our clusters are detected from the optical CFHTLS data described above in Section \ref{lsdat}. The overdensity of red-sequence galaxies was used as tracers of clusters \citep{GY00}. The detection procedure was described in full detail in  \cite{lu09}. From the 171 square degrees, we detected  $\sim$200 clusters with $N_{red,m^*+2}\geq 12$ in the redshift range $0.16<z<0.36$, where  $N_{red,m^*+2}$ is the number of red-sequence galaxies brighter  than $m^*+2$, within a radius of 0.5 Mpc from the cluster centre. The richness  $N_{red,m^*+2} \sim 12$ roughly corresponds to a mass of $10^{14}$ $\som$ (see \citealt{lu09} for details). In Figure \ref{richdis} we plot the richness vs. redshift of the clusters in our sample. Because the GALEX data are only available for about 65 per cent of the whole CFHTLS fields, for our analysis we use a subset of 112 clusters with GALEX coverage.  To do the background subtraction as we will describe in the next Section, we divide our sample into two redshift bins, $0.16<z<0.27$ and $0.28<z<0.36$, and stack the clusters in each bin to increase the statistics. The number of clusters in the two bins are 43 and 69 respectively.

\subsection{Background Subtraction}\label{mul-sub}

To study the properties of galaxies in clusters, it is important to separate them from the foreground/background galaxies that are projected along the line-of-sight to the clusters of interest. We use the photometric redshift catalogue of the CFHTLS Wide survey \citep{ilbert06,Coupon2009} to reduce the contamination by removing galaxies whose 2$\sigma$ redshift uncertainties do not put them within $\Delta z = \pm 0.1$ around the redshift of interest. We reduce the contamination further by using the photometric redshift catalogue to outline the region in the multi-colour space occupied by galaxies at the redshift of interest, and exclude galaxies (including those without photometric redshift) that do not fall within the 99 per cent contour. 

Note that the redshift distribution of cluster members is much narrower than the typical uncertainty of the photometric redshift. Therefore, a statistical background subtraction is still needed, especially when we examine the outskirt regions of clusters where the contrast with the field is low. Somewhat different from the usual background subtraction method used in previous studies, we take advantage of our multi-band photometry and  do the subtraction in multi-colour space, as we will describe in detail below. Our technique allows the removal of objects which do not have plausible colours to fit any typical SED at the redshift of interest. This will be important when we turn to models to convert observed quantities to physical ones. This  multi-colour subtraction technique is somewhat analogous to fitting empirical photometric redshifts  (e.g., \citealt{bolzonella00}). 

To construct the fore/background contamination sample (referred to as the background  sample from now on), we do exactly the same as with the cluster sample, but replace the centre of each cluster with a random position generated from the same patch (W1,2,3,4) the cluster was detected from\footnote{For the background sample, we mask out any known clusters, out to $r=3$ Mpc, though this makes no appreciable difference.}. This way, for each cluster in the stack, there is a corresponding background sample with the same systematics such as the imaging depth and overall number density (which could vary from patch to patch due to different Galactic extinction and cosmic variance). 

Due to the masking (around bright stars and regions with elevated background as discussed in Sections \ref{lsdat} and \ref{gdat}) and the gaps between GALEX tiles, the areal data coverage for each cluster is not 100 per cent. Therefore, we need to calculate, for each cluster, within the radius of interest, what fraction of the area is covered by valid data. Only clusters with a valid coverage greater than 80 per cent are included. We estimate the coverage fraction by using a high-resolution random catalogue, where points are randomly distributed over the footprint of our survey with an average density of 1 object per 10 square arcsec (a balance between computing efficiency and resolution requirement). We then weight each galaxy by this fraction, and by the number of clusters in the stack so that the number attached to each galaxy represents the number per cluster, i.e.
\begin{eqnarray}
 w_i= 1\Big\slash\left(\frac{N_{data}}{N_{rand}}\ n_{cl}\right),\label{wt}
\end{eqnarray}
 where $N_{data}/N_{rand}$ is the areal fraction covered by valid data and $n_{cl}$ is the number of clusters in the stack.

With the two sets of samples in hand, one with cluster + background counts, and one with just background counts, the background-subtracted net count of each galaxy, $n_{i}$, is simply:
\begin{eqnarray}
 n_{i}= w_{cl+bg,i} -w_{bg,i},
\end{eqnarray}
where $w_{bg,i}$ is the weight (as in Equation \ref{wt}) of a background galaxy, and $w_{cl+bg,i}$ is the weight of the nearest cluster+background counterpart to that background galaxy. The nearest counterpart is defined to be the galaxy in the cluster+background sample that is closest to that background galaxy in the space of $NUV$ magnitude, $(NUV-u^*),\ (u^*-g'),\ (g'-r'),\ (r'-i')$, and $(i'-z')$ colours with equal weight\footnote{It makes no significant difference to the results whether or not we incorporate the colour uncertainties in this procedure.}.

If the weight of the background galaxy, $w_{bg,i}$, is greater than that of its counterpart, $w_{cl+bg,i}$, the resulting $n_{i}$ would be negative. In that case, we set $n_{i}$ to zero, and  subtract the ``negative excess" from  the next-nearest counterpart, and so on until the net excess becomes positive. In the end, we  have a subset of galaxies from the cluster+background sample that all have a positive number attached to them. These galaxies should essentially all be cluster members, because galaxies that have similar spectral shape as the background galaxies are effectively removed by doing the subtraction in multi-colour space. This way, not only is the total magnitude distribution of cluster members recovered \footnote{We have verified this by comparing the magnitude distribution of the cluster members obtained from our subtraction method and the more traditional subtraction method.}, but colour-dependent quantities such as star formation rate can be derived.

We perform the last two steps, constructing the background sample and subtracting it from the cluster sample, 100 times, and average the results over the 100 realizations. The error bars on the results presented throughout the paper are the standard deviation of the 100 realizations. Thus, they reflect the field-to-field variance in the background, which is the dominant source of statistical error in our measurements.

\begin{figure*} 
 \includegraphics[width=\textwidth]{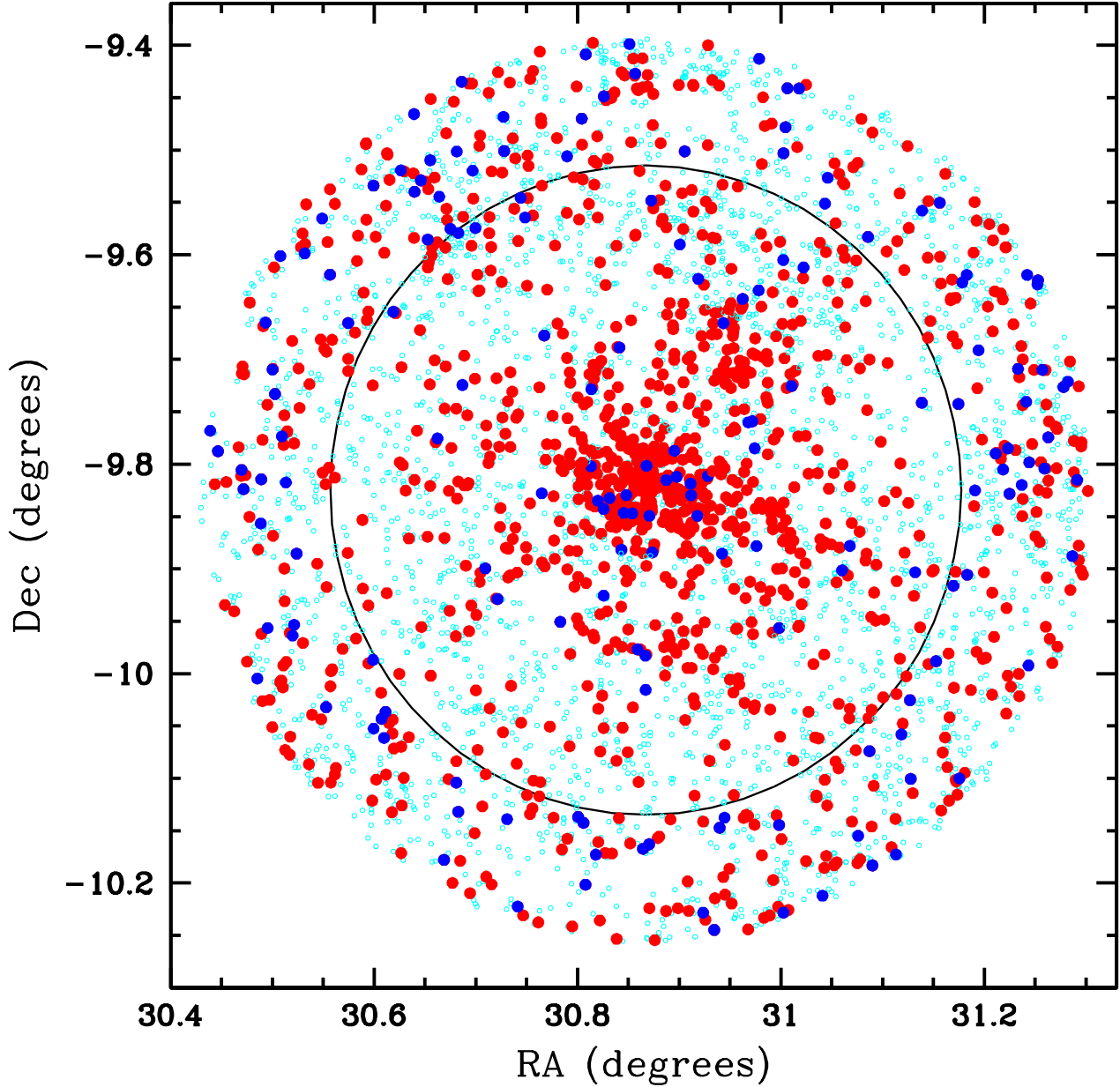}   \caption[Spatial distribution of galaxies in the field of view (r$<$7Mpc) of one example cluster]{The spatial distribution of galaxies in the field of view (r$<$7Mpc) of one example cluster. Open cyan circles are all potential cluster members, as determined from their photometric redshifts. Solid red dots are galaxies that are left after the background subtraction, with NUV detected sources indicated in blue. The big black circle shows the 5 Mpc radius. \label{eg_sp}}
\end{figure*}
\begin{figure*} 
 \includegraphics[width=0.5\textwidth]{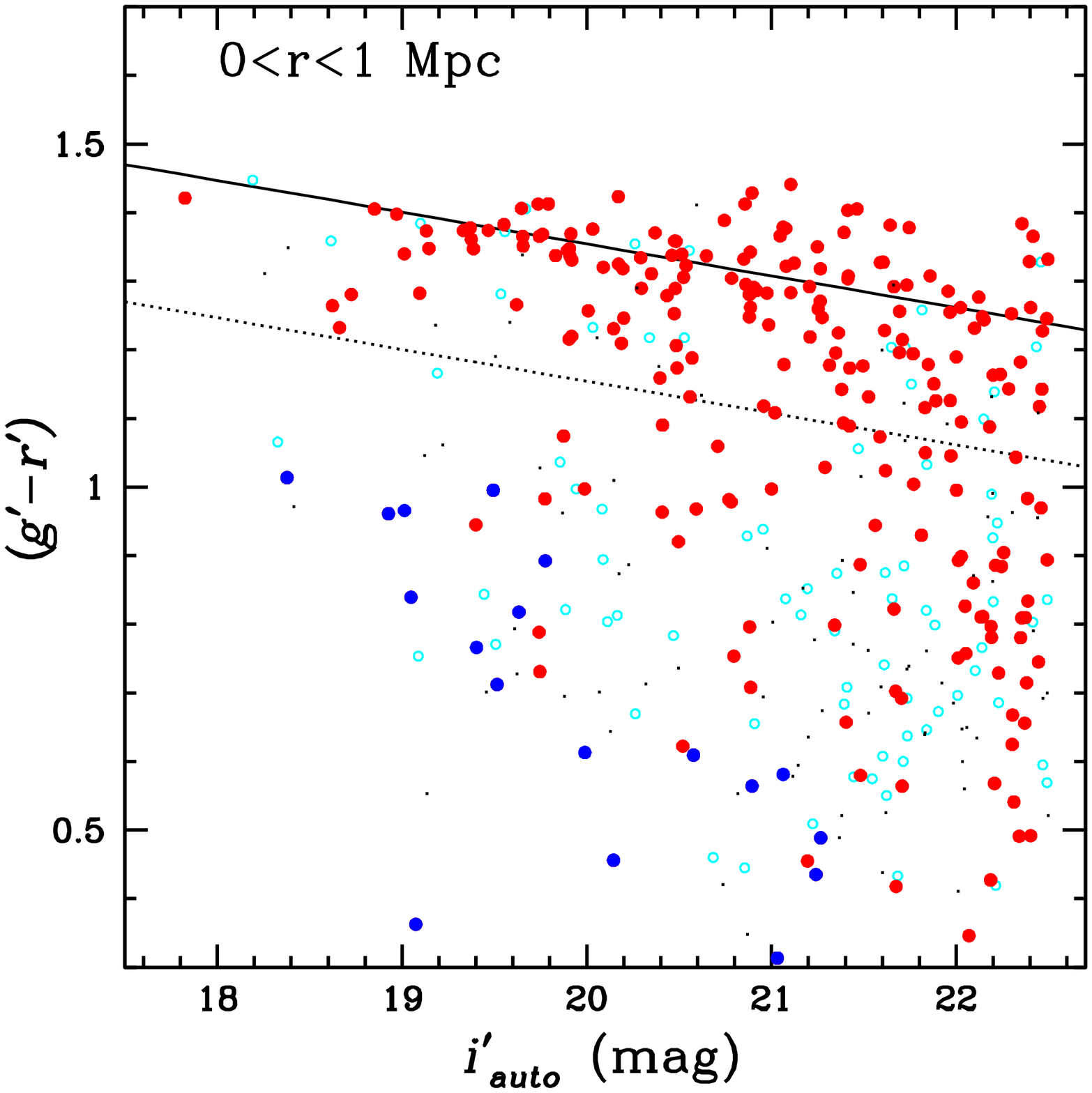}\includegraphics[width=0.5\textwidth]{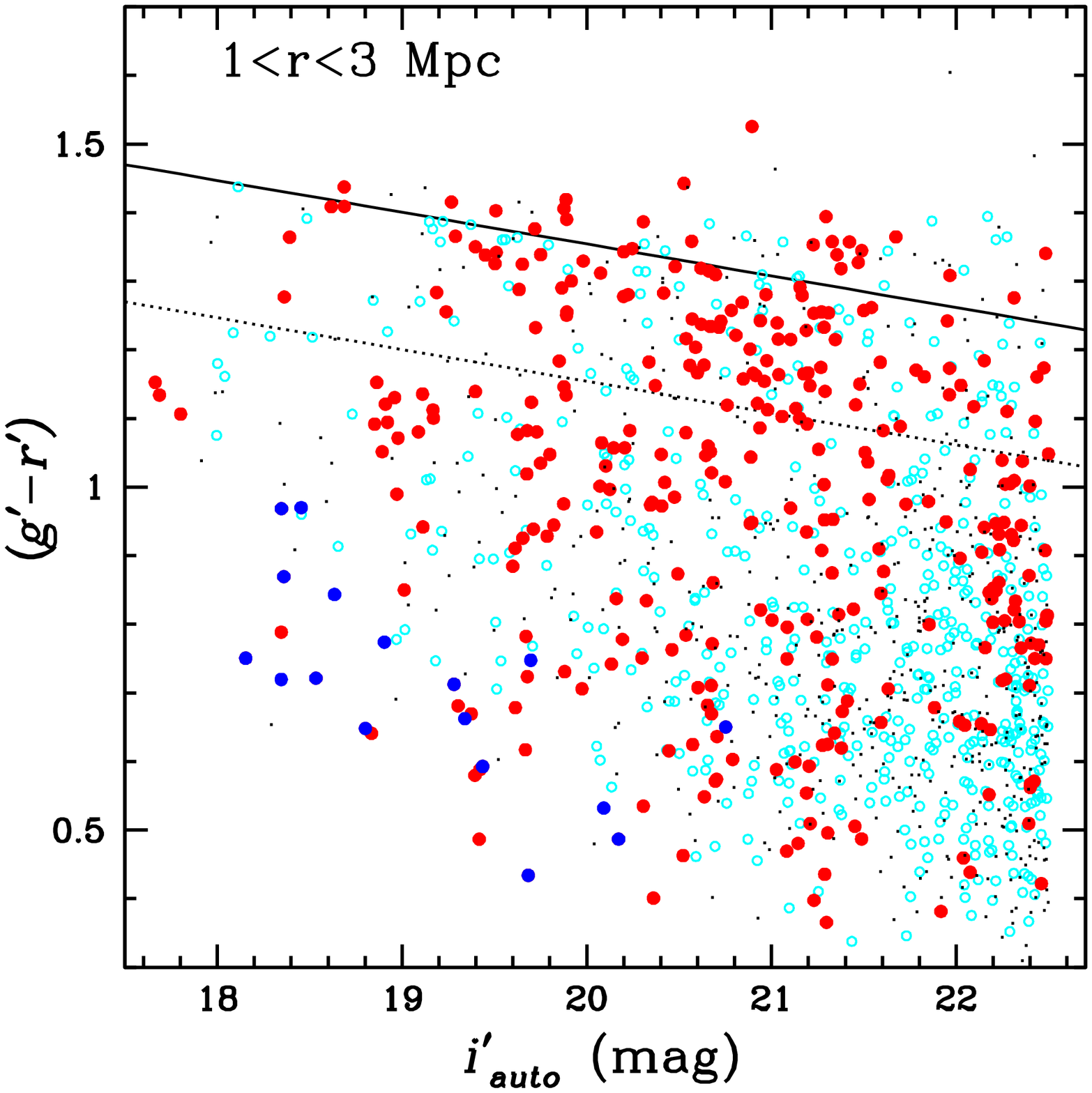}    
 \includegraphics[width=0.5\textwidth]{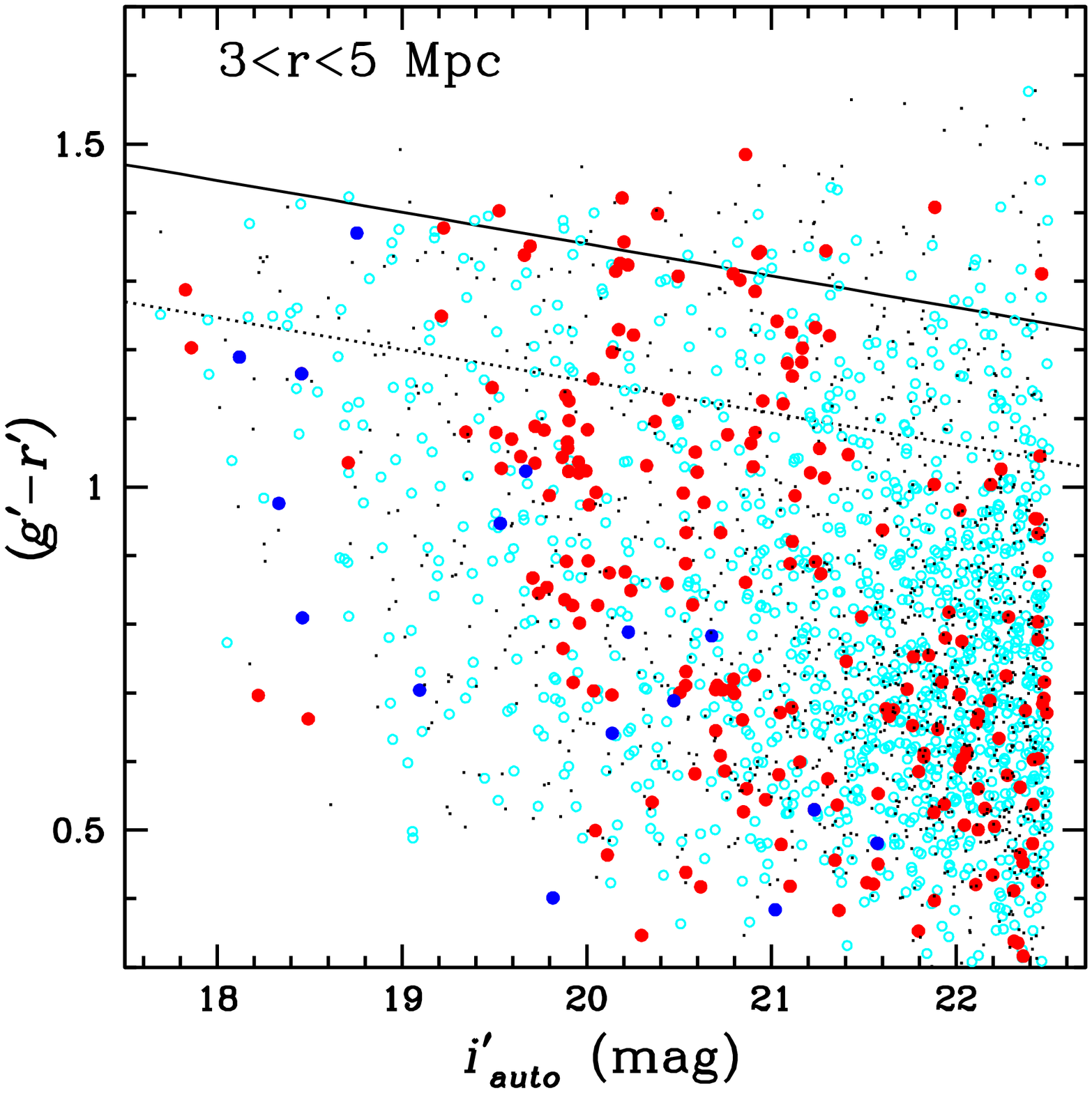}\includegraphics[width=0.5\textwidth]{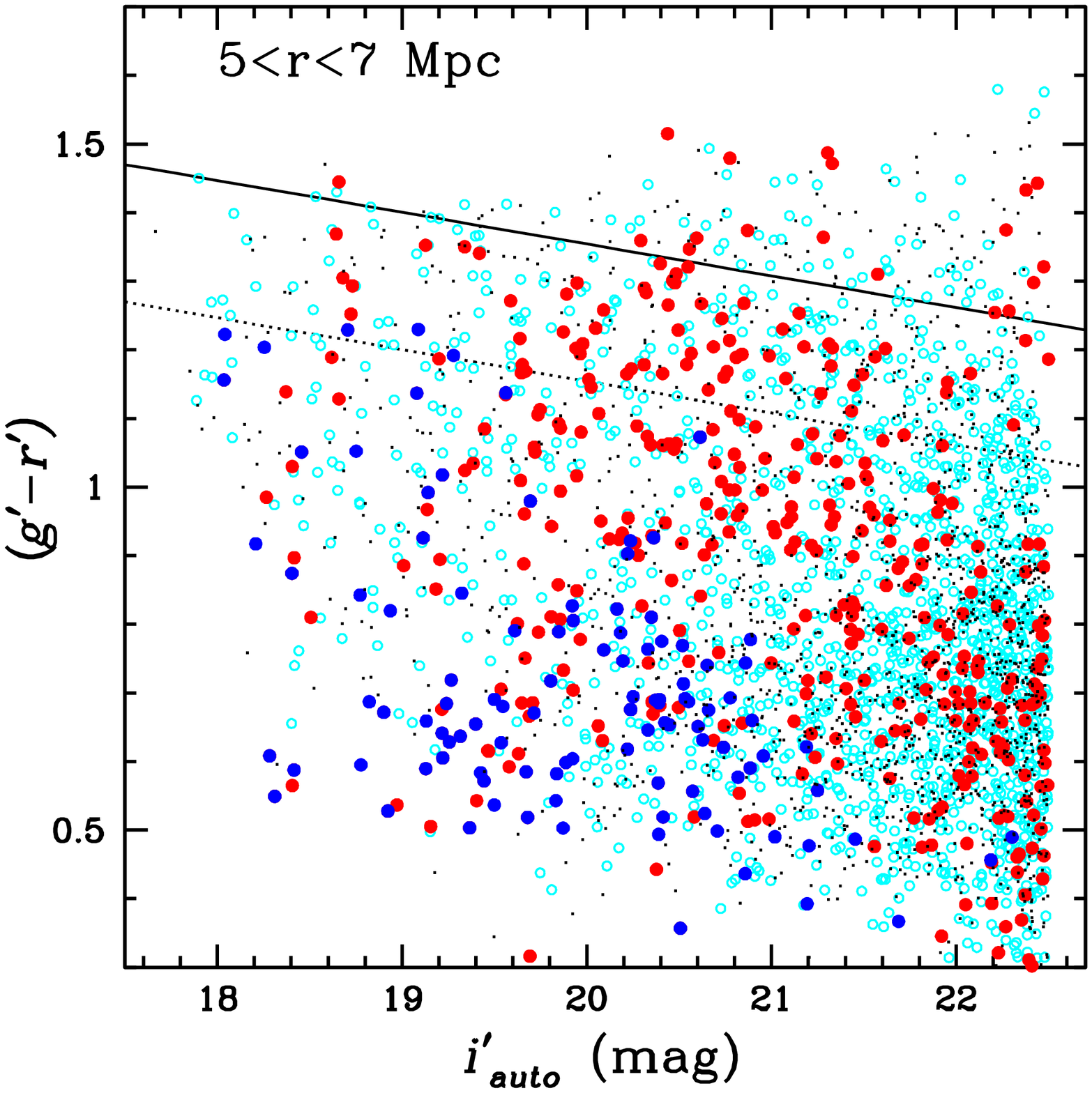}    \caption[CMDs of the example cluster]{The colour-magnitude diagrams (CMDs) of the example cluster shown in Figure \ref{eg_sp}, in four radial bins. The symbols are the same as in Figure \ref{eg_sp}, with the additional black points indicating background galaxies. The solid lines are the model CMD from \cite{galexev}. To guide the eye, the dashed line indicate colours 0.2 mag bluer than the red-sequence.  \label{eg_gri}}
\end{figure*}

As an example, in Figure \ref{eg_sp} we show the result of one realization of the background subtraction of one of the clusters in our sample. Open cyan circles represent all potential cluster members in the field of view before the subtraction, with photometric redshift pre-selection. Solid red dots are galaxies that are left after the subtraction, with UV detected galaxies indicated by the blue dots. The large black circle shows a radius of 5 Mpc from the cluster centre. The colour-magnitude diagrams (CMDs) of this cluster, in four radial bins, are shown in Figure \ref{eg_gri}. The symbols are the same as in Figure \ref{eg_sp}, with the additional black points indicating background galaxies. The red-sequence is clearly visible in the core of the cluster.

\subsection{Phot-z Field Sample}

\begin{figure}
\includegraphics[width=0.5\textwidth]{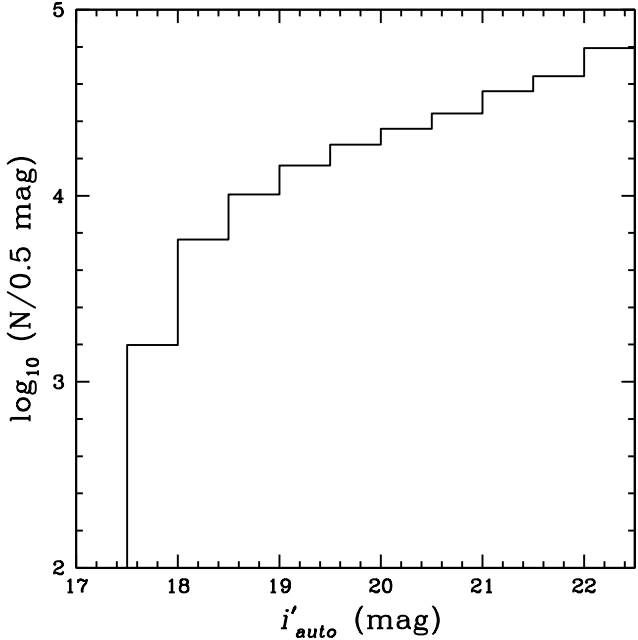}\caption{Number of galaxies in the field sample, as a function of $i'_{auto}$ magnitude. Because of the large sample size, the statistical uncertainties on the field values in our analysis are negligible.}\label{fd_appi}
\end{figure}

To  compare the properties of galaxies in clusters with that in the general field, we construct a comparison field sample. We select, from the parent catalogue where clusters are detected, galaxies in the redshift range covered by our cluster sample using their phot-z. For the field sample, the  phot-z of each galaxy is used in the calculation of the $k$-correction (Section \ref{kcor}). In Figure \ref{fd_appi} we show the number of galaxies in the field sample as a function of  $i'_{auto}$ magnitude. Because of the large sample size, the statistical uncertainties on the field values in our analysis are negligible and thus are not plotted in the Figures shown in this paper.

\section{Derivation of Physical Quantities}\label{anal}

\subsection{k-corrections}\label{kcor}
After the background subtraction what we are left with are cluster members. But note, these members come from a number of clusters in a relatively wide redshift bin, stacked together. By doing so, the original redshift of the host cluster (determined from the red-sequence galaxies as described in detail in \citealt{lu09}) is no longer relevant. Therefore, for galaxies within each redshift bin, the median redshift of the bin is used when computing the $k$-correction. The $k$-corrections are computed using {\sc kcorrect} v4\_1\_4 \citep{Bkcor}. Because at the redshift of our cluster sample, the observed NUV magnitude probes the FUV wavelength, we reconstruct the rest-frame FUV magnitude from our six passband photometry using the public code of \cite{Bkcor}. The typical correction is about 0.6 mag.

\subsection{Stellar Mass}
\begin{figure} 
 \includegraphics[width=0.5\textwidth]{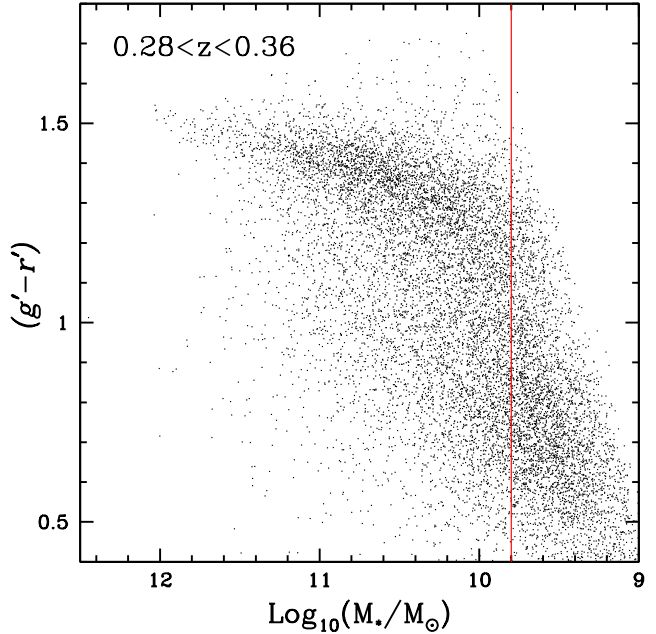}
\caption[Stellar mass completeness.]{The colour vs. stellar mass plot for galaxies in our higher redshift cluster sample. The mass-to-light ratio is a function of $(g'-r')$, thus for the red galaxies we are complete down to  $10^{9.8}$ $\som$, as indicated by the vertical solid red line.   \label{stmlim}}
\end{figure}

We estimate stellar mass using the $i$-band mass-to-light ratio, as a function of SDSS $(g-r)$ colour, from  \cite{BellML}. We  project our k-corrected rest-frame $(g'-r')$ colour measured in CFHTLS filters onto the SDSS system, using the code of \cite{Bkcor}. Note that this conversion is provided for a diet Salpeter IMF, and therefore to be consistent with the Kroupa IMF used in the UV SFR estimation, the resulting stellar mass is scaled by 0.7 \citep{BellML} to convert to the standard Salpeter IMF and then divided by 1.5 \citep{brinchmann04} to convert to Kroupa IMF. Because the mass-to-light ratio we used is only a function of colour, we have a well defined stellar mass completeness as a function of colour. In Figure \ref{stmlim} we show for our higher redshift cluster sample, the colour vs. stellar mass. It shows that for red galaxies we are complete down to  $10^{9.8}$ $\som$, as indicated by the vertical solid red line. We only consider galaxies above this stellar mass in the analysis.

\subsection{Star Formation Rate}\label{sfr}
\begin{figure*} 
 \includegraphics[width=0.46\textwidth]{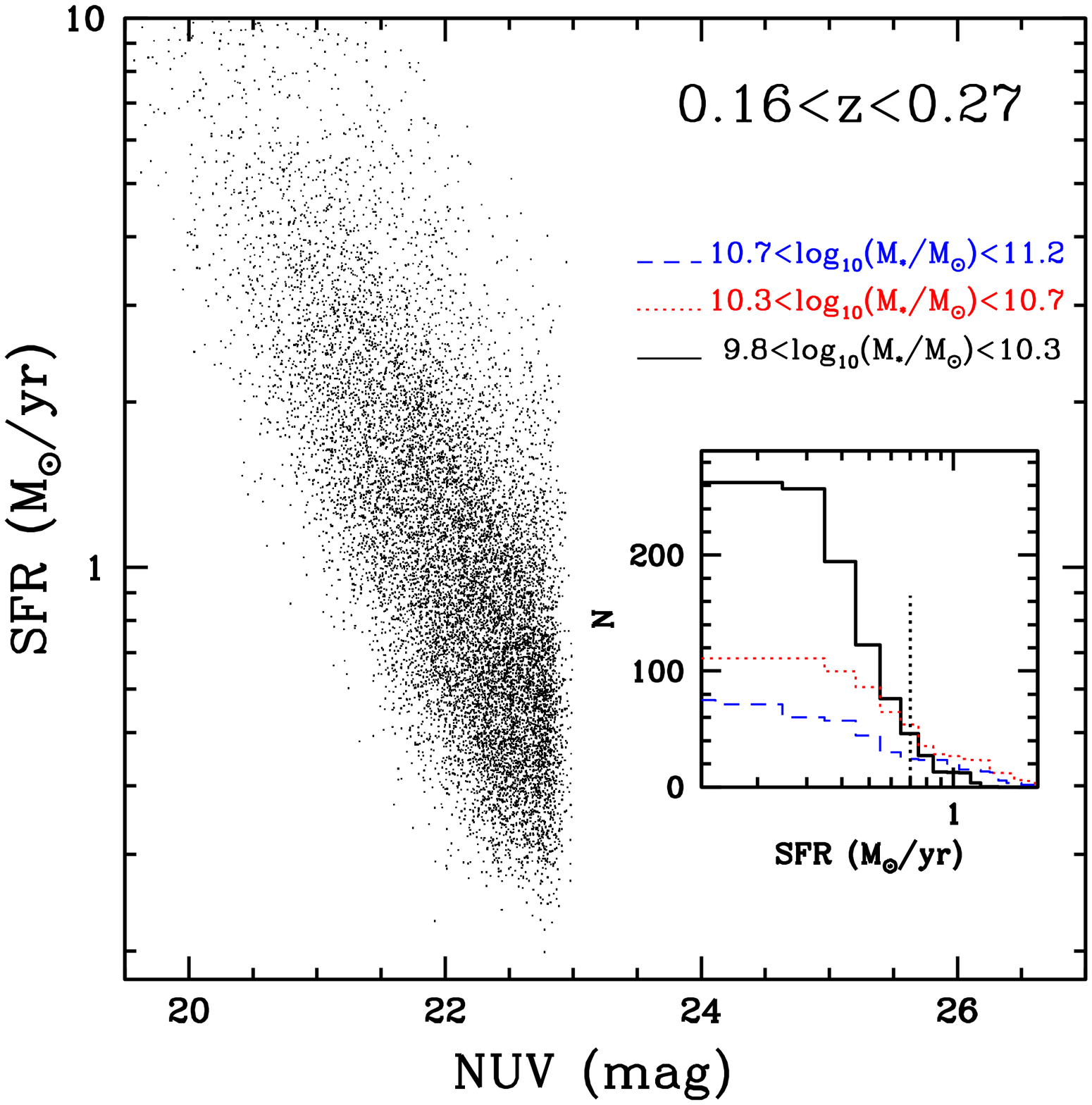} \includegraphics[width=0.46\textwidth]{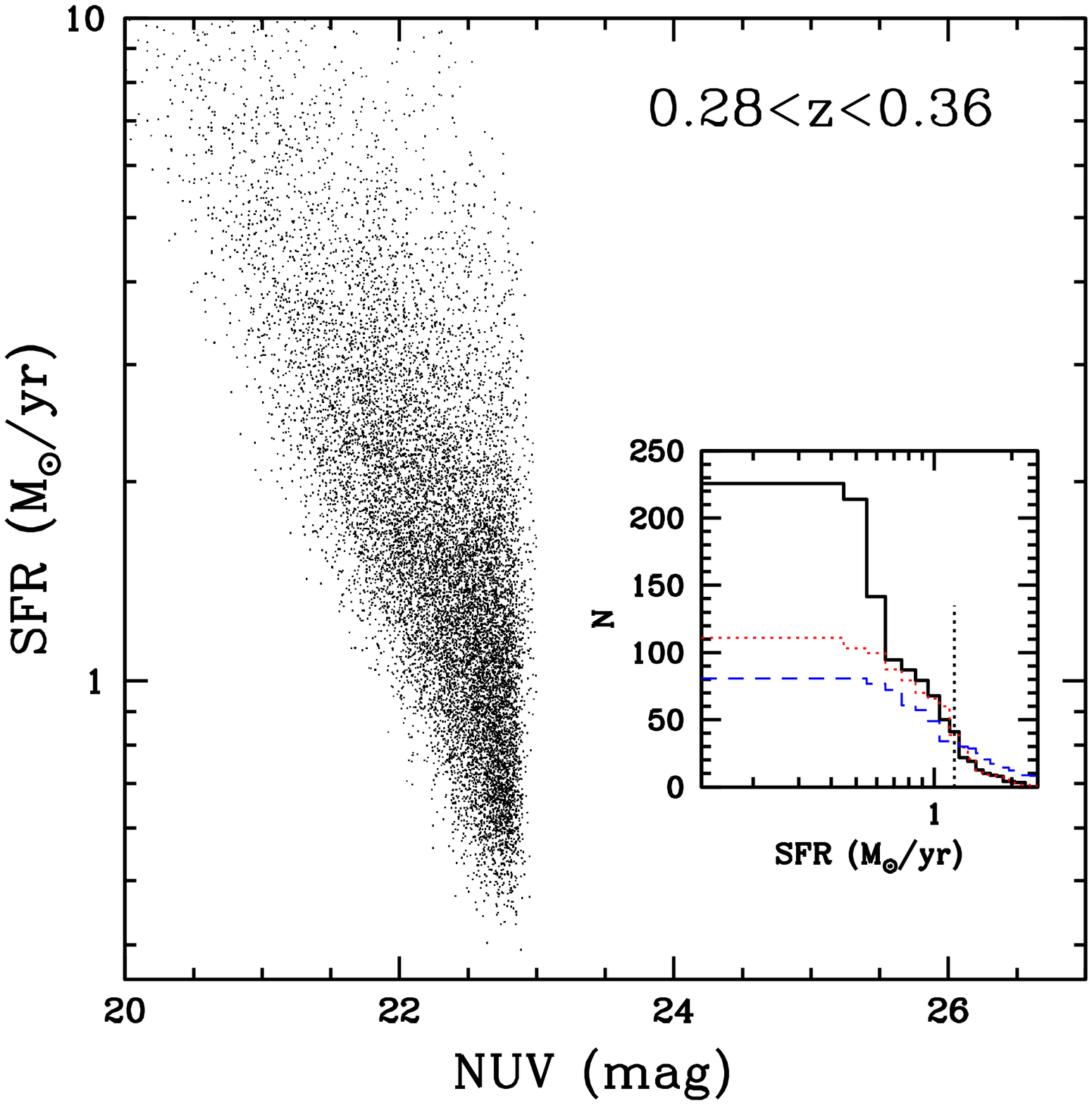}
\caption{SFR vs. apparent NUV magnitude, at the two redshifts respectively. The insets show the cumulative SFR distribution of galaxies $\sim$0.2 magnitude brighter than the apparent NUV magnitude limit (NUV=23 mag). Different histograms are for different stellar masses, and the vertical lines indicate the adopted SFR$_{\text{UV}}$ limits for our analysis. Depending on the stellar mass, $\sim 60-90$ per cent of the galaxies near the NUV detection limit is below the SFR limits we adopt (0.7 $\som$/yr at $z\sim 0.2$ and 1.2 $\som$/yr at $z\sim 0.3$).\label{sfrlim}}
\end{figure*}

\begin{figure*} 
 \includegraphics[width=0.46\textwidth]{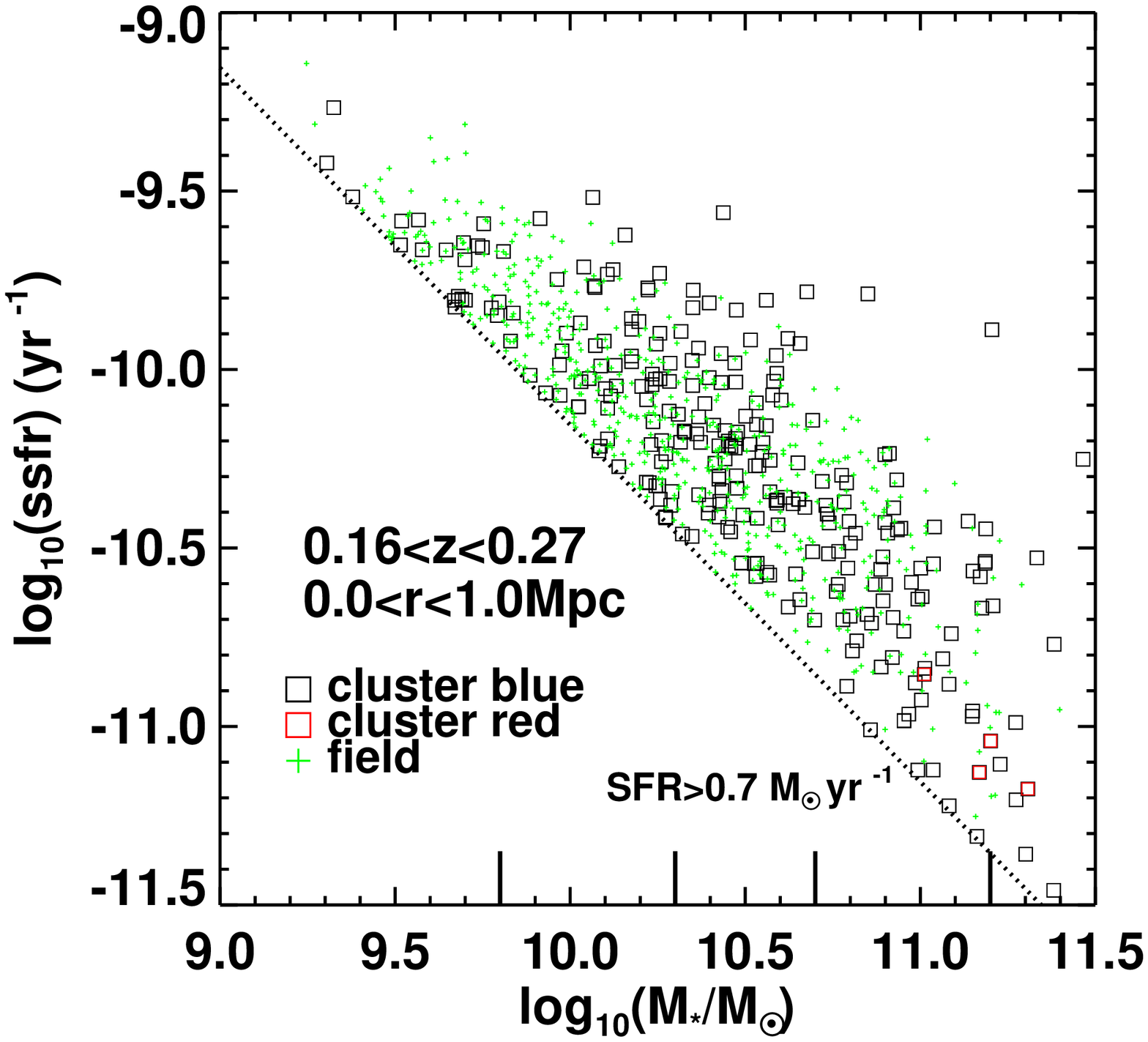}\includegraphics[width=0.46\textwidth]{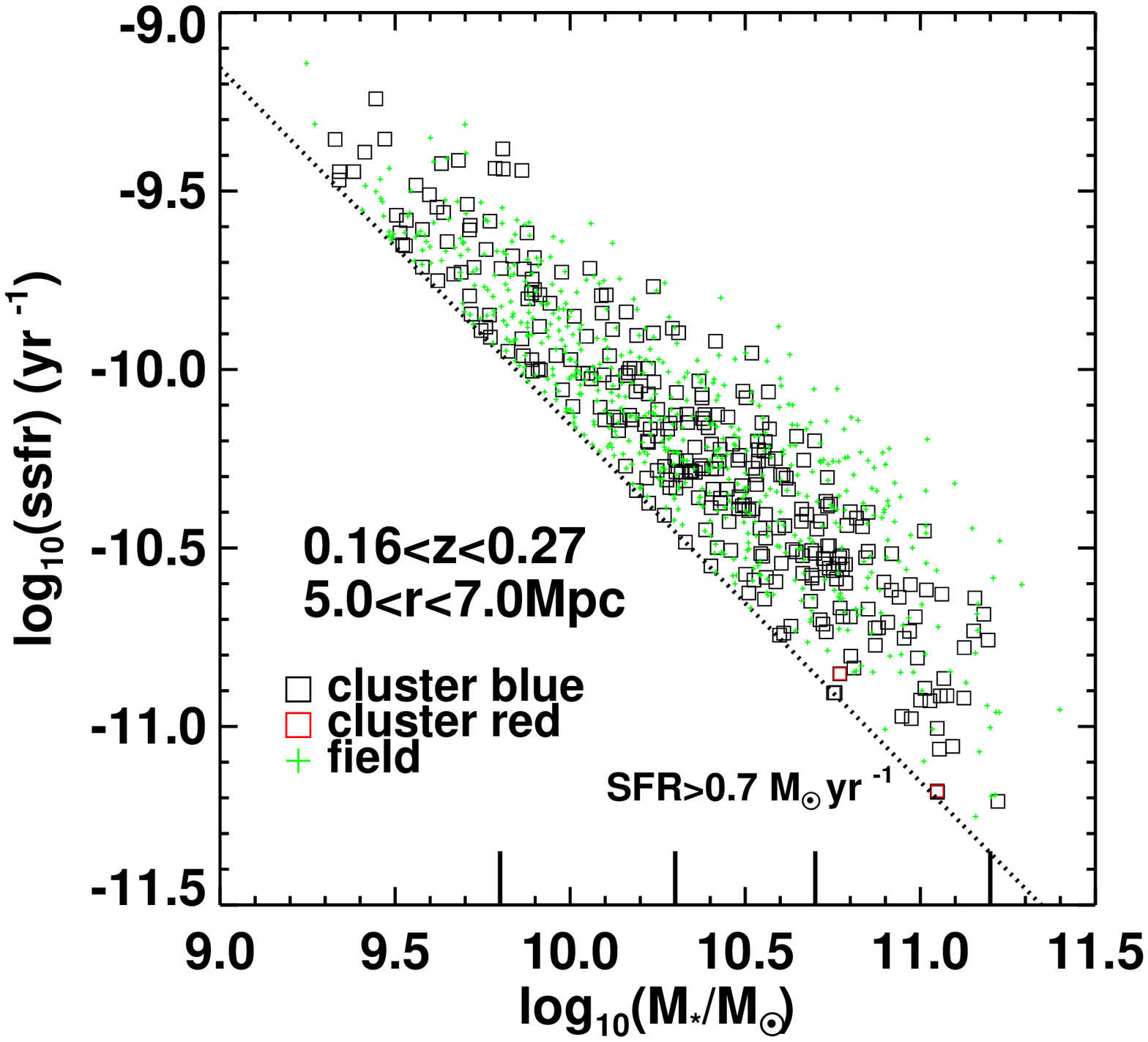}
\includegraphics[width=0.46\textwidth]{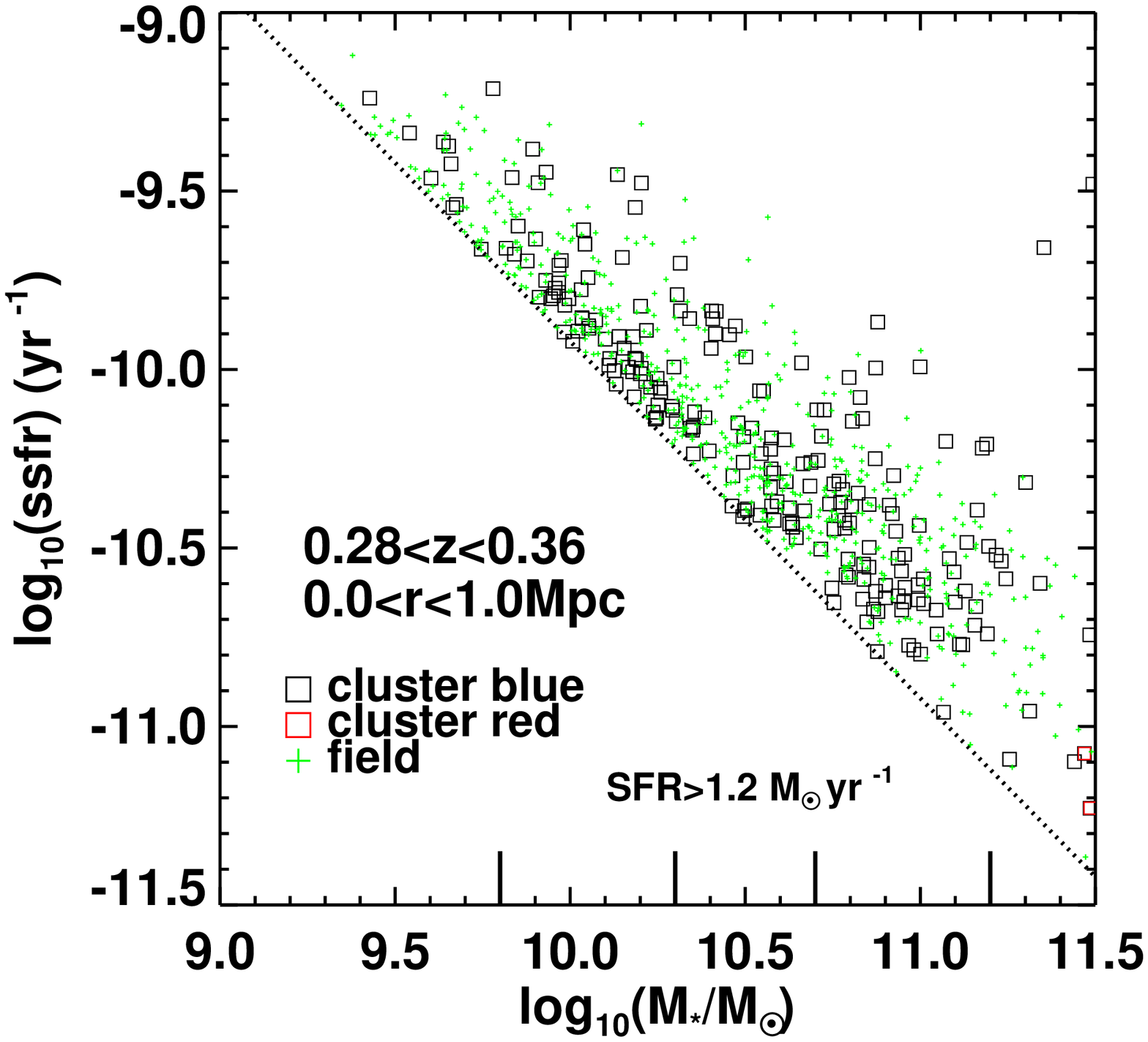}\includegraphics[width=0.46\textwidth]{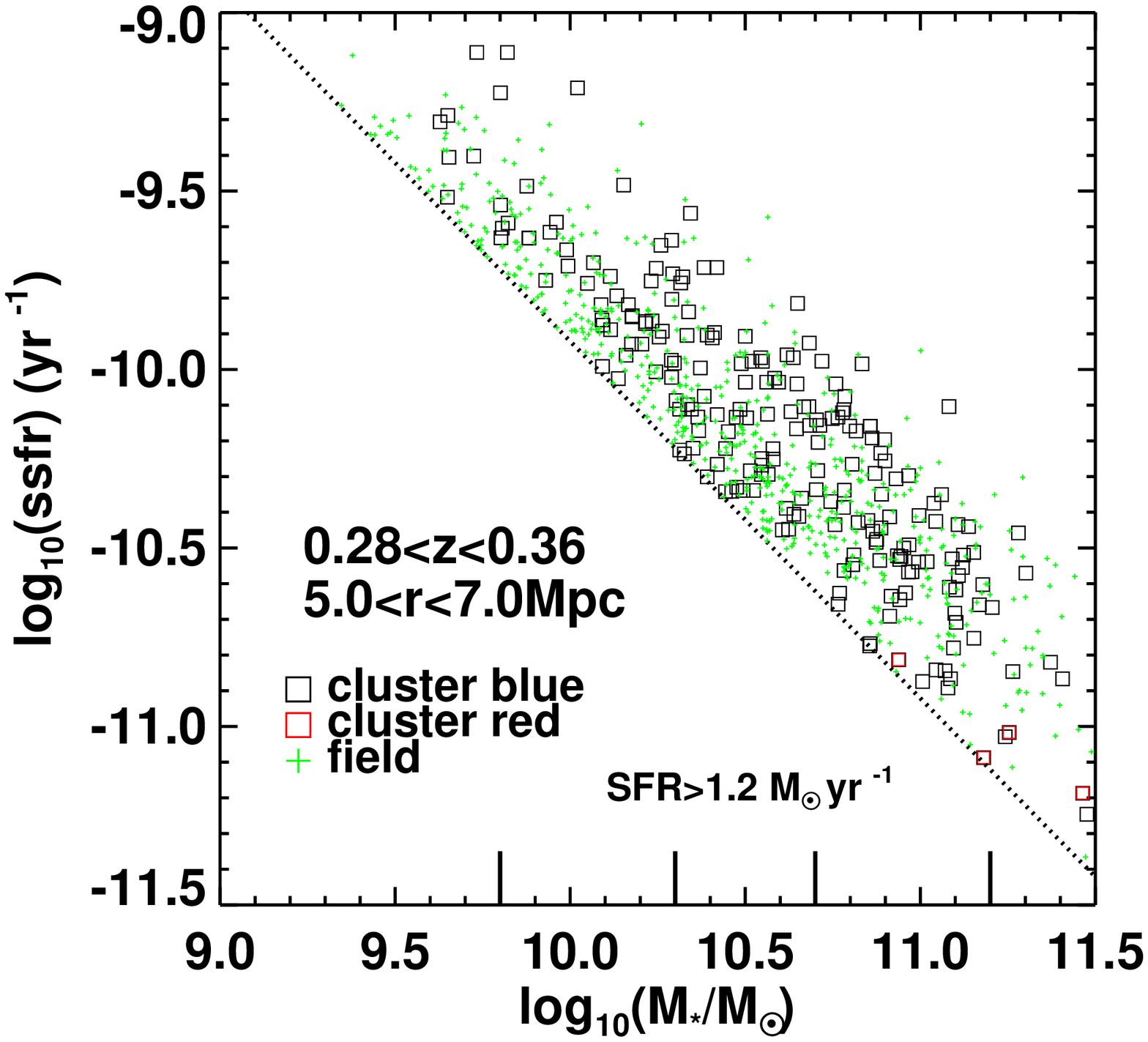}
\caption[SSFR vs. stellar mass of our stacked clusters]{One realization of the  SSFR vs. stellar mass relation of our stacked clusters at two redshifts (top and bottom panels), for the two regions with the most contrasting density, the core ($0<r<1$Mpc) (left panels) and the outskirt ($5<r<7$Mpc) (right panels).  The black squares are blue cluster galaxies, and the red squares represent those that are on the red-sequence. Galaxies in the phot-z field sample are plotted as green crosses. The  slanted dashed lines indicate the UV detection limits. The short solid lines indicate the three stellar mass bins used in the analysis. \label{ssfrm}}
\end{figure*}

\begin{figure*} 
 \includegraphics[width=0.47\textwidth]{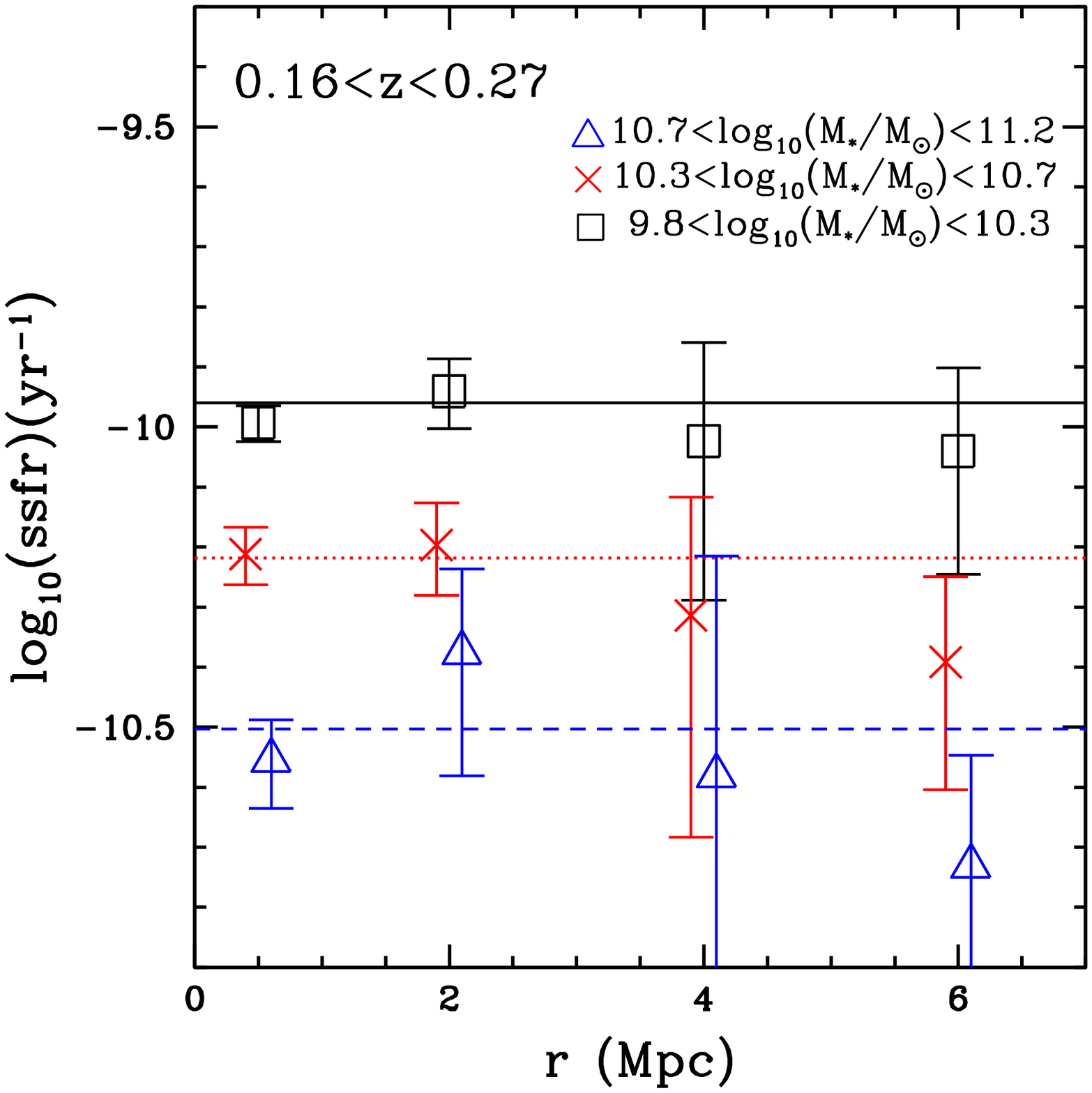} 
\includegraphics[width=0.47\textwidth]{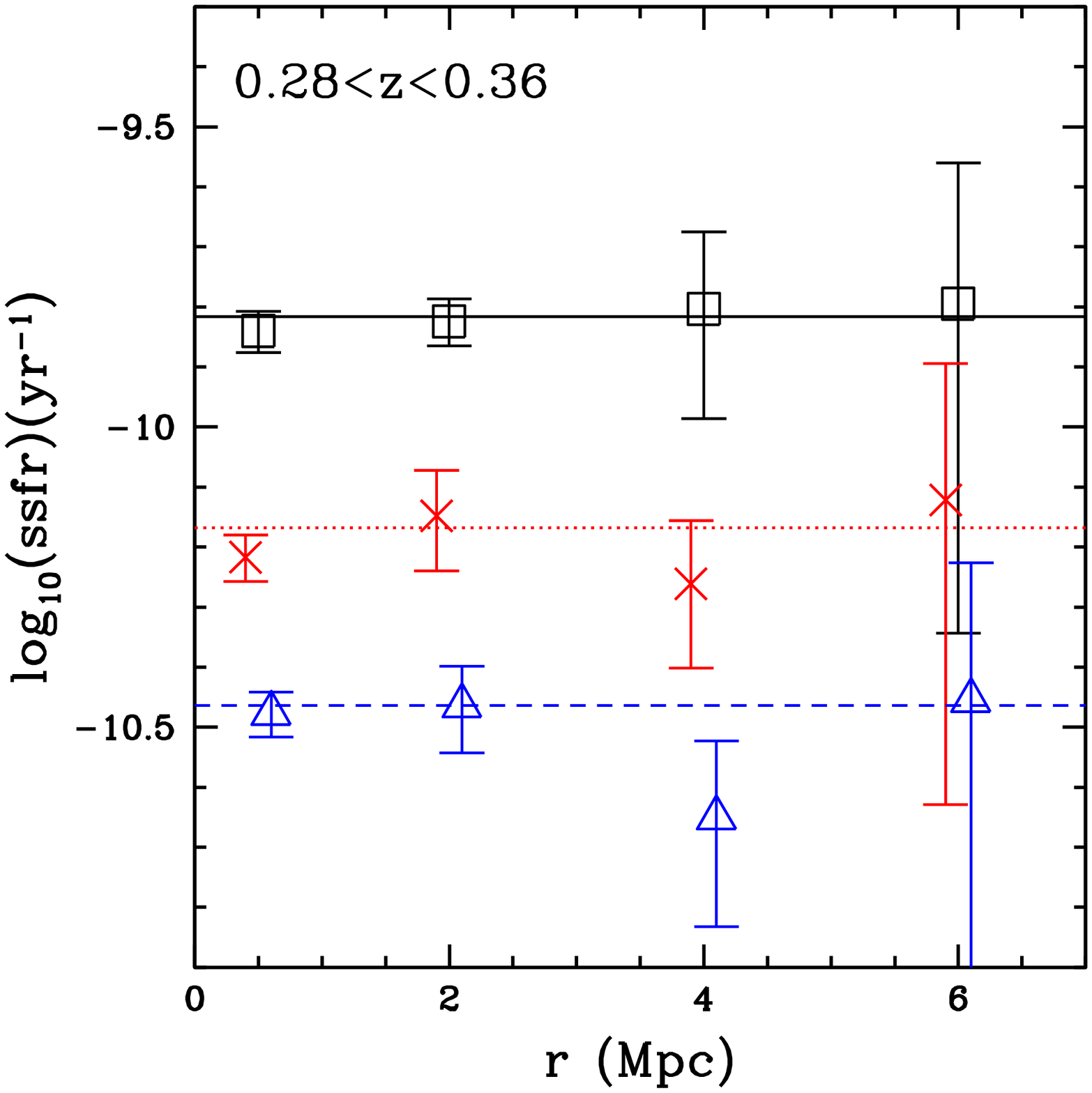} \caption[Ensemble average SSFR  vs. r]{The ensemble average SSFR  (total SFR/total stellar mass) of galaxies with SFR above our UV detection limit, at different cluster-centric radii, split into three stellar mass bins. Black squares, red crosses, and blue triangles represent the $9.8<\text{log}_{10}(\text{M}_*/\som) < 10.3$, $10.3<\text{log}_{10}(\text{M}_*/\som) < 10.7$, and $10.7<\text{log}_{10}(\text{M}_*/\som) < 11.2$ mass bins, respectively. The corresponding field values are indicated by the black solid, red dotted, and blue dashed lines respectively. The average SFR  of the star-forming galaxies (above our SFR$_{\text{UV}}$ limits) is roughly constant from the cluster core out to the outskirt regions, and is consistent with the field value within the uncertainties. \label{avgssfr}}
\end{figure*}

The UV luminosity we observe cannot be directly converted to a star formation rate because of the existence of dust   around star-forming regions. The dust will absorb the UV emission and re-emit it at a longer wavelength. Therefore, it is important to correct for the effect of dust absorption, before we can infer the star formation rate from the UV luminosity.

The total IR  to UV luminosity ratio ($L_{TIR}/L_{UV}$) is a reliable estimator of the dust extinction in star-forming galaxies   \citep[e.g.][]{buat92}. We do not have IR information for our cluster sample. However, it has been shown that there is a correlation between $L_{TIR}/L_{UV}$ and the slope of the UV continuum \citep[e.g.][]{meurer99,cort06}. We use this to  estimate $L_{TIR}/L_{UV}$, and eventually the dust extinction.

If fit by a power law, the UV continuum can be described as $f(\lambda)\propto \lambda ^{\beta} $, where $f$ is in unit of $ergs\ s^{-1}\ cm^{-2}\ \lambda^{-1}$ \citep{cal94}. Therefore, the slope $\beta$ can be estimated as $\beta=\left({{\log}f_2-{\log}f_1}\right)/\left({{\log}\lambda_2-{\log}\lambda_1}\right)$. In the redshift range of our cluster sample, the rest-frame wavelength range used to measure the UV slope, 1250\AA$<\lambda < 2600$\AA,  is shifted  to the bandpasses of NUV and $u^*$. Therefore, in our case, we reconstruct the rest-frame FUV and NUV from our observed multi-band photometry again using the code of \cite{Bkcor}. 
 Converting the AB magnitude to flux using $m(AB)=-2.5{\log}f_{\nu}-48.6=-2.5{\log}(f_{\lambda} \lambda^2/c)-48.6$, where $f_{\nu}$ is in unit of ergs~s$^{-1}$~cm$^{-2}$~Hz$^{-1}$, we have:\\ 
$\beta=-0.4\left({m_{u^*}-m_{NUV}}\right)/\left({{\log}\lambda_2-{\log}\lambda_1}\right)-2=2.2(m_{FUV}-m_{NUV})-2$, where $m_{FUV}$ and $m_{NUV}$ are in the rest-frame.

It has been pointed out that the relation between the UV slope and  $L_{TIR}/L_{UV}$ is different for star-bursting galaxies and normal star-forming galaxies \citep{bell02,kong04,cort06}. Since most of the galaxies we are after here are normal star-forming galaxies, we adopt the relation derived by \cite{cort06}(eq. 5) for normal star-forming galaxies, which is: 
\begin{equation}
{\log}(L_{TIR}/L_{FUV})=(0.7\pm0.06)\beta+(1.3\pm 0.06).
\end{equation}

In the study by  \cite{cort08}, the relation between  $L_{TIR}/L_{FUV}$ and dust extinction A(FUV) is derived  using SED fitting for galaxies with different ages. Therefore, for galaxies off the red-sequence (defined as those with $(g'-r')$ and $(r'-i')$ colour bluer by 0.2 magnitude than the red-sequence), we adopt the relation for  young star-forming galaxies (their $\tau>7$ Gyr model); and for galaxies on the red-sequence we adopt the one for an older population (their $\tau\sim 5.4$ Gyr model) (see their table 1). As suggested by \cite{cort08}, for galaxies redder than $(FUV-NUV)=0.9$, their extinction is assumed to be the same as galaxies with $(FUV-NUV)=0.9$.

The dust-corrected magnitude is  turned into a luminosity, and the star formation rate is then estimated using the  \citet{kennicutt} relation. The factor of 1.5 \citep{brinchmann04} is to convert from  Salpeter IMF \citep{salpeter} to Kroupa IMF \citep{kroupa01}: 
\begin{equation}
 SFR_{\text{UV}}(\som \mbox{/yr})=1.4\times 10^{-28} L_{\nu} (\mbox{ergs}\ \mbox{s}^{-1}\ \mbox{Hz}^{-1})\times 10^{0.4 A_{\tau}}/1.5.
\end{equation}
 
Because of the above described conversion between NUV magnitude and SFR, there is a scatter between the observed NUV magnitude and the derived SFR, as shown in Figure \ref{sfrlim} (small dots). Thus, the limiting NUV magnitude of our sample does not correspond to a precisely defined SFR limit. We can estimate the SFR limit by looking at the SFR of galaxies near the limiting apparent NUV magnitude. The insets in Figure \ref{sfrlim} show  the cumulative SFR distribution of galaxies $\sim$0.2 magnitude brighter than the NUV magnitude limit (NUV=23 mag) at the two redshifts respectively. Different histograms are for different stellar masses. At  $z\sim 0.2$, depending on the stellar mass,  $\sim 60-90$ per cent of these galaxies near the NUV detection limit have SFR below 0.7 $\som$/yr. Therefore, we adopt 0.7 $\som$/yr as the SFR detection limit for our NUV magnitude limited sample at this redshift. Similarly, at  $z\sim 0.3$, we adopt a SFR limit of 1.2 $\som$/yr. These limits are indicated as the vertical lines in the insets. The use of a different limiting SFR at each redshift bin means we must be cautious when interpreting trend with redshift as evolution. Our main purpose, however, is to compare the field and cluster populations at each redshift.

\section{Results}\label{gres}

In this section, we present our findings on the specific star formation rate (SSFR), fraction of star-forming and blue/red galaxies of the cluster sample as a function of cluster-centric radius, and compare them with that of the field sample. Note that, as discussed before, we divide our sample into two redshift bins, and thus the limiting SFR will be different for each bin.

\subsection{SSFR vs. Stellar Mass}\label{sfrdis}

\begin{figure*} 
 \includegraphics[width=0.47\textwidth]{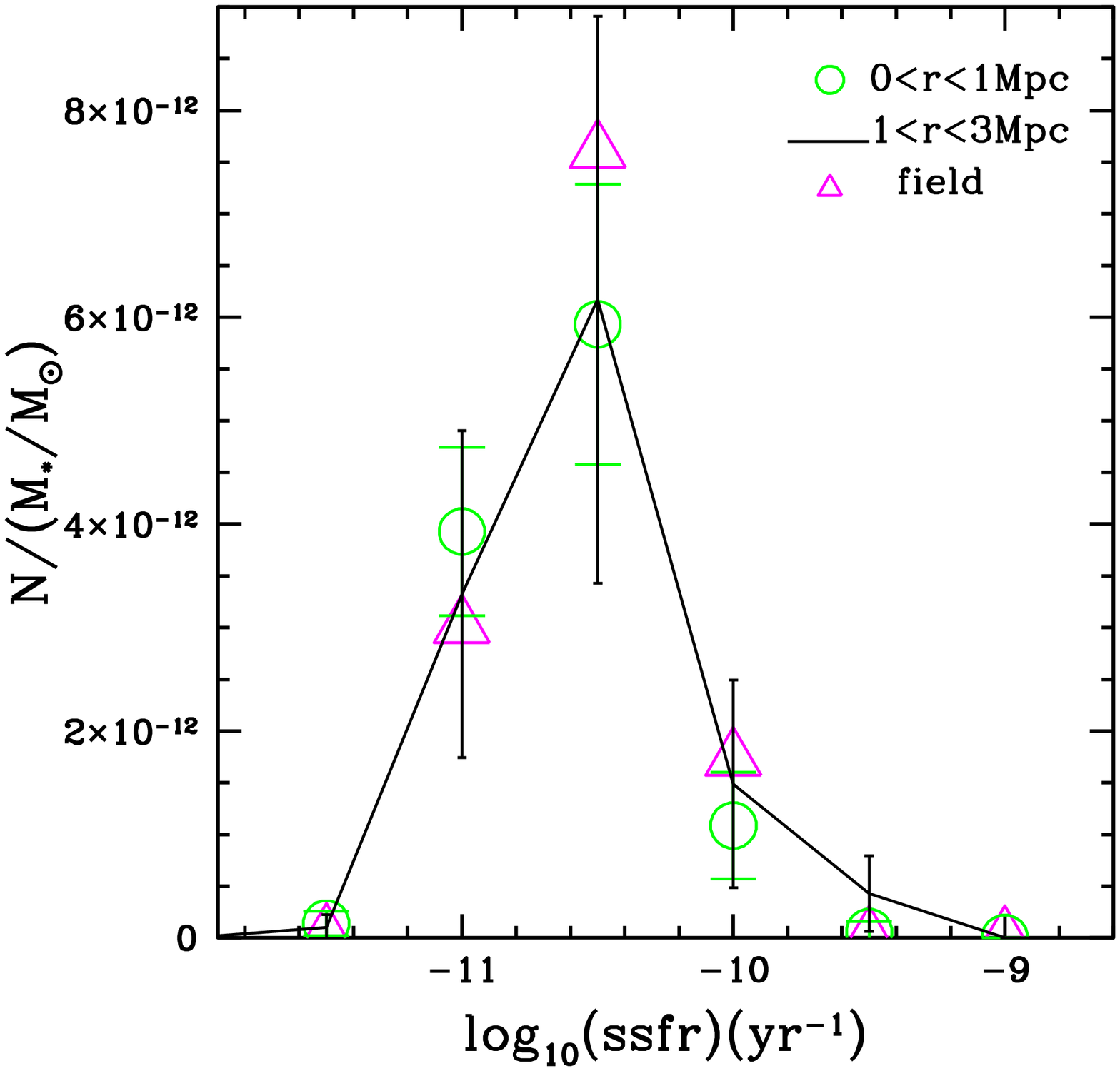}\includegraphics[width=0.47\textwidth]{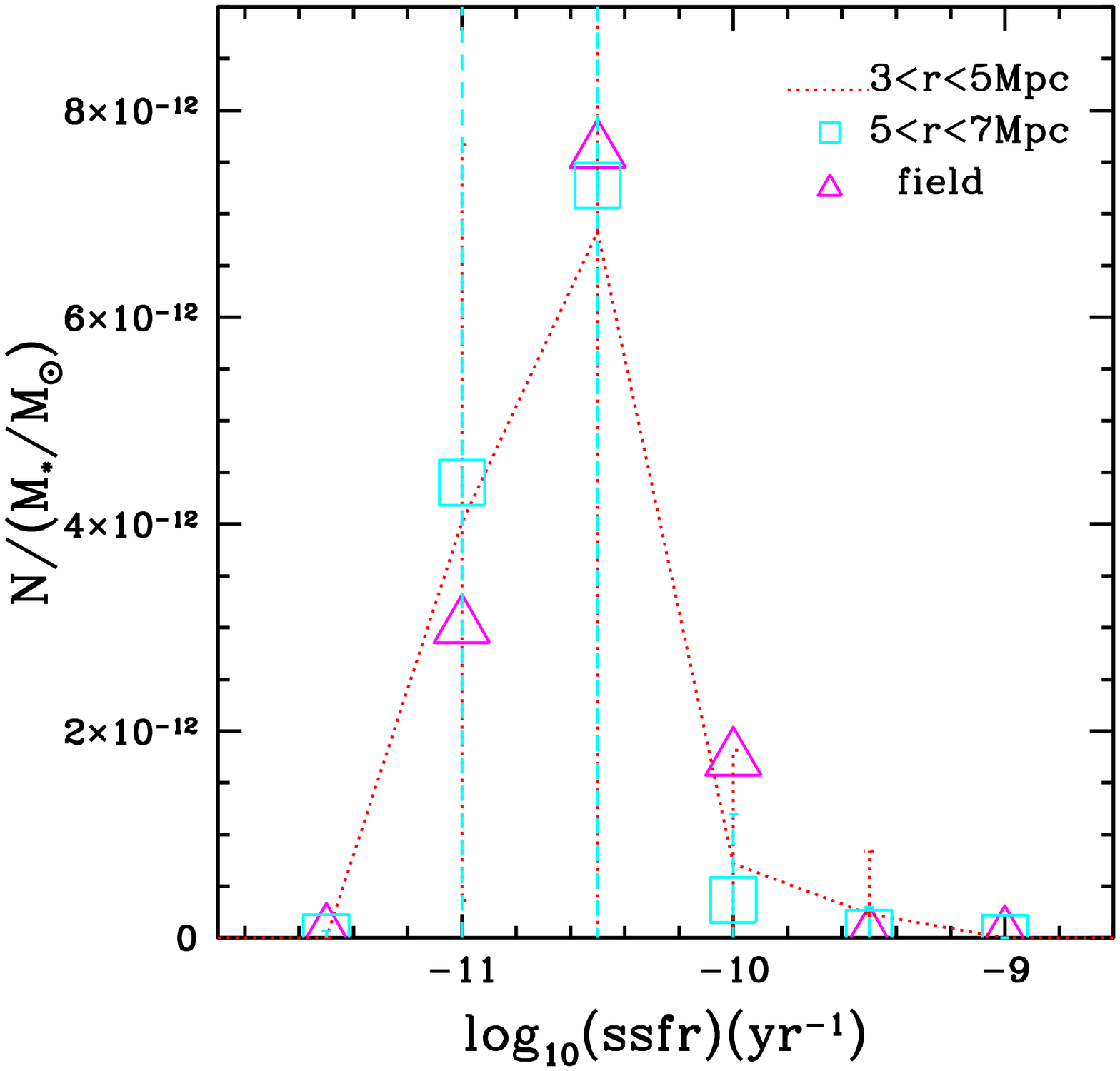}
 \caption[Distribution of the SSFR]{The distribution of the SSFR of galaxies with  $10.7<\text{log}_{10}(\text{M}_*/\som) < 11.2$ at $0.16<z<0.27$, normalized by the total stellar mass, at different radii. The green circles, black solid and red dotted histograms, and cyan squares represent the four cluster-centric radii, and magenta triangles represent the field values. For clarity, we split it into two panels with the field values plotted in both as references. Within the relatively large error bars, there does not seem to be a significant difference between the distribution at different cluster-centric radii and the field.  \label{ssfrdis}}
\end{figure*}

It has been suggested that star-forming galaxies form a tight sequence in the SSFR vs. stellar mass (M$_*$) plane \citep{noeske07,Salim2007} (but see e.g. \citealt{Cowie2008} where a much larger scatter is found), where SSFR is defined as SFR/M$_*$.  In Figure \ref{ssfrm}, we show one realization of the  SSFR vs.  M$_*$ of our stacked clusters at two redshifts, in the top and bottom panels respectively. The black squares are blue cluster galaxies, while red squares represent those that are on the red-sequence. The green crosses are phot-z field galaxies. In the left and right panels, we present the relation in two regions with the most contrasting density: the core ($0<r<1$Mpc) and the outskirts ($5<r<7$Mpc). Note that our SFR$_{\text{UV}}$ detection limits (as discussed in Section \ref{sfr}) are 0.7 $\som$ yr$^{-1}$ at $z\sim 0.2$  and 1.2 $\som$ yr$^{-1}$ at $z\sim0.3$ (as indicated by the slanted line). Therefore, the slope of the SSFR vs. M$_*$ of our sample is mostly driven by the detection limit, and does not reflect an intrinsic correlation between these quantities. For this reason, we do not attempt to fit the relation for our sample. Nonetheless, comparing the left panels with the right ones, it seems that the relation in the cluster cores and the outskirts are both  consistent with that in the field, showing no dependence on the environment. 

The ensemble average SSFR  (total SFR/total stellar mass) of galaxies with SFR above our UV detection limit is shown in Figure \ref{avgssfr}, at different cluster-centric radii, split into three stellar mass bins. Black squares, red crosses, and blue triangles represent the $9.8<\text{log}_{10}(\text{M}_*/\som) < 10.3$, $10.3<\text{log}_{10}(\text{M}_*/\som) < 10.7$, and $10.7<\text{log}_{10}(\text{M}_*/\som) < 11.2$ mass bins respectively. For all stellar masses,  the average SSFR of the star-forming galaxies (above our adopted SFR$_{\text{UV}}$ limits) is approximately constant from the cluster core out to the outskirt region. The Spearman's rank correlation tests show that the observed relation between the average SSFR and the radius deviates from no-correlation hypothesis on $1-1.5\ \sigma$ level for the various stellar masses. Note that, our errorbars prevent the accurate determination of the rankings of the variables; if ranking the average SSFR randomly, the level of deviation varies from $\sim 0.3-1.7\ \sigma$. 

It is also interesting to compare this with the field values, which are indicated by the black solid, red dotted, and blue dashed lines for the three stellar masses respectively. As mentioned before the statistical uncertainties on the field values are entirely negligible, given the large sample size. Within the uncertainties, there does not appear to be a significant difference between the average SSFR of star-forming galaxies in clusters and  in the field.

We now further examine the distribution of the SSFR (instead of just the average) in each stellar mass bin. For clarity, we only show the one for the most massive galaxies ($10.7<\text{log}_{10}(\text{M}_*/\som) < 11.2$) in the lower redshift bin ($0.16<z<0.26$), as the behavior of galaxies at other stellar masses and at the higher redshift is qualitatively the same. In Figure \ref{ssfrdis}, the green circles, black solid and red dotted histograms,  cyan squares, and magenta triangles represent respectively the distribution (normalized by the total stellar mass) of the SSFR at four cluster-centric radii and in the field. For clarity, we split them into two panels with the field values (magenta triangles) plotted in both panels as references. The distributions appear to be consistent within our relatively large error bars. This implies that, above our SFR$_{\text{UV}}$ detection limits, either the quenching timescale is short  so that the population of transition galaxies eludes detection, or that only a small fraction of the population above the limits are going through  the transition and the majority of the star-forming galaxies are left alone.

There results are in agreement with \cite{peng10,mcgalex10,biv11,wetzel11} in general, but in apparent conflict with the low SSFR found in clusters by \cite{vul2010}. We will discuss this further in Section \ref{gdis}.

\subsection{Star-forming Fraction}\label{s_fsf}
We now examine  the fraction of star-forming galaxies out of the whole population as a function of environment.
To obtain the total number of galaxies, we follow the same procedure as for the optical-UV matched catalogue,  except that we now include galaxies that are not NUV detected. 

\begin{figure*} 
 \includegraphics[width=0.47\textwidth]{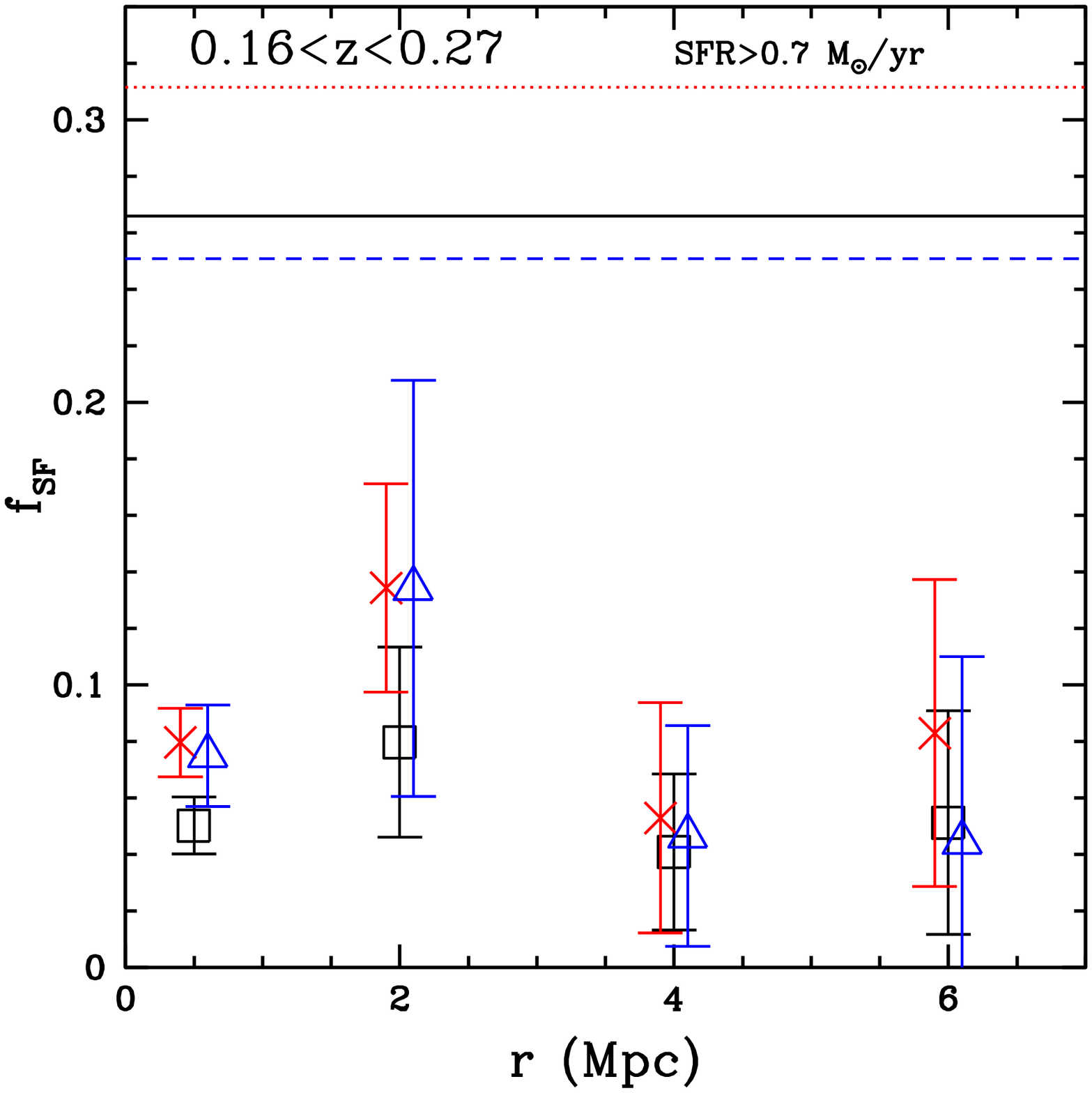}
\includegraphics[width=0.47\textwidth]{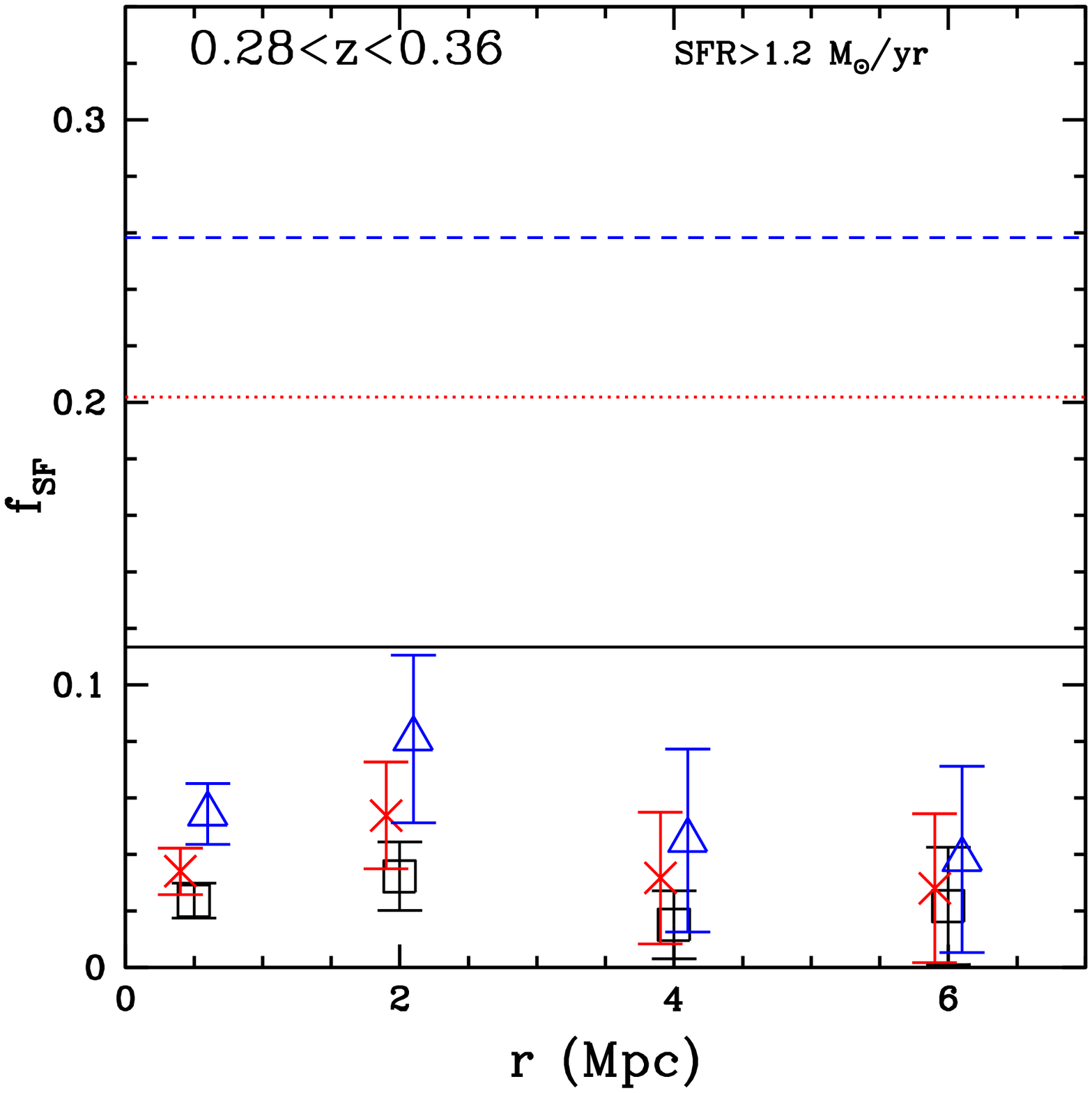}  
\caption[Fraction of star-forming galaxies]{The fraction of star-forming galaxies with SFR above our UV detection limit, as a function of cluster-centric radius in three stellar mass bins. Blue triangles, red crosses, and black squares represent three stellar mass bins (in decreasing order, as labeled in Figure \ref{fred}), and the corresponding field values are indicated by blue dashed, red dotted, and black solid lines. The lower and higher redshift samples are shown in the left and right panels respectively. \label{fsf}}
\end{figure*}

In Figure \ref{fsf}, the fraction of star-forming galaxies with SFR above our UV detection limit is plotted as a function of cluster-centric radius in three stellar mass bins, for the lower (left panel) and higher redshift samples (right panel). Symbols are for cluster galaxies at three stellar masses (as labeled in Figure \ref{fred}), and the lines are corresponding field values. For all stellar masses at both redshifts, the fraction of star-forming galaxies in clusters is much lower than that in the field, and is independent of the radius within the uncertainties. The very low star-forming fraction in clusters is partially due to our relatively high SFR$_{\text{UV}}$ detection limits.  Note that, the stellar mass dependence of the star-forming fraction shown here (for both clusters and the field) may seem counterintuitive, but keep in mind that our star-forming fraction is calculated with fixed cuts in SFR  instead of SSFR. Therefore, although the star-forming fraction is higher for low mass galaxies, their typical SFR is lower and thus more of them would be below our SFR cut. In any case, our focus here is the comparison between clusters and field in each redshift bin, not the stellar mass dependence.

In the next section, we examine the behavior of the blue population that includes galaxies with SFR lower than our UV detection limit.

\subsection{Blue/Red Fraction}\label{s_fred}

\begin{figure*} 
\includegraphics[width=0.47\textwidth]{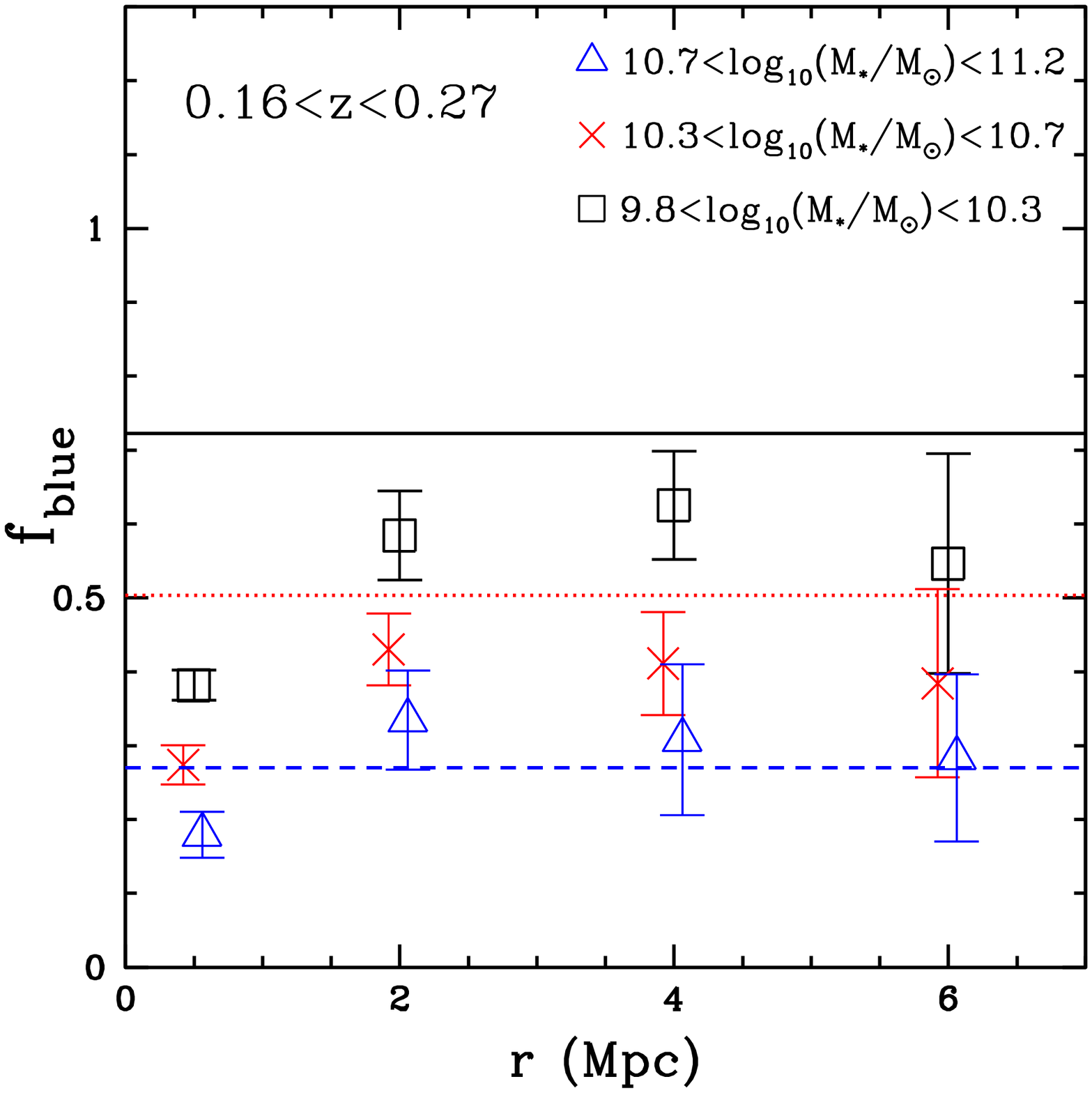}
\includegraphics[width=0.47\textwidth]{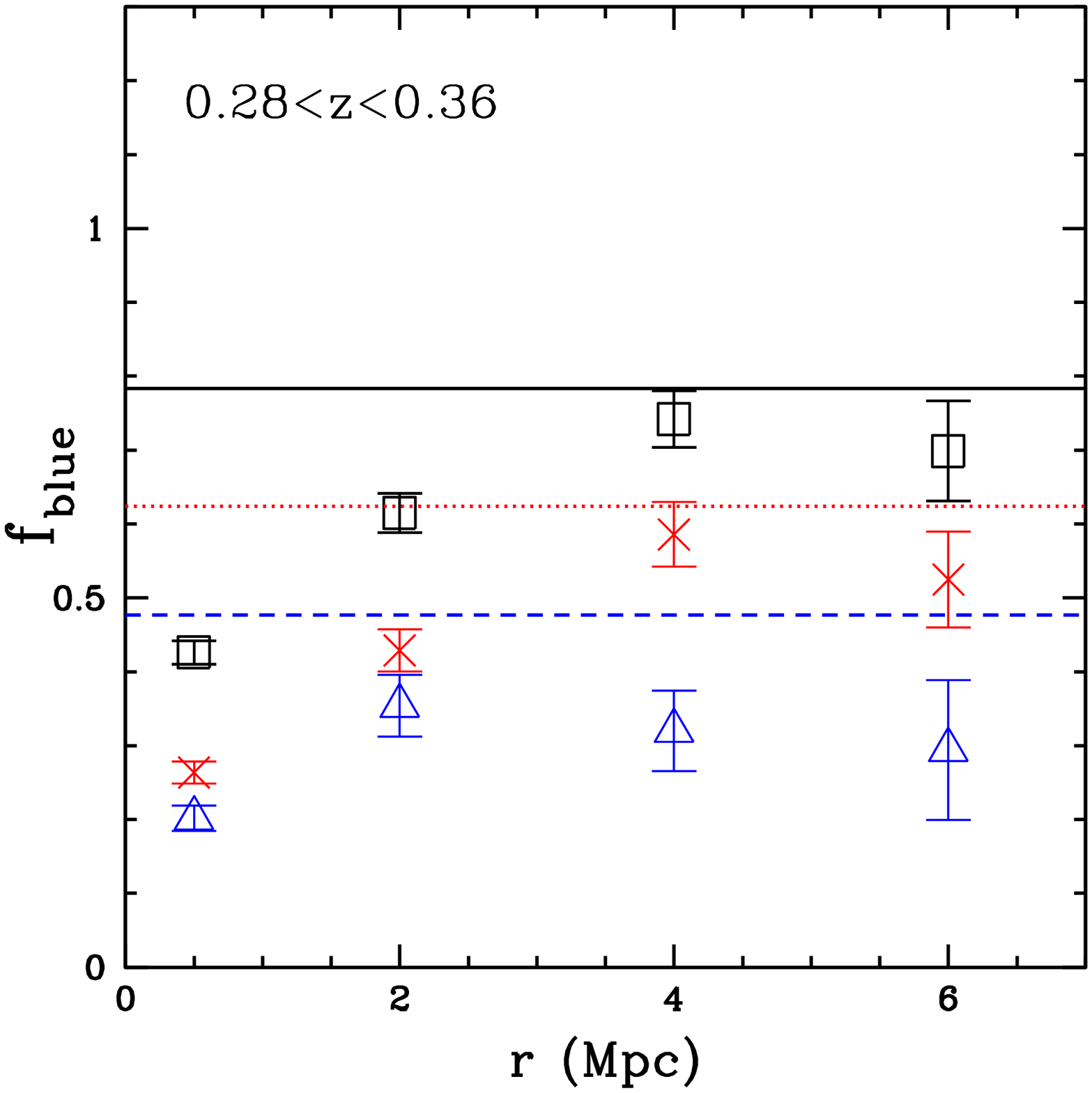}   
\caption[Fraction of blue galaxies]{ The fraction of blue galaxies as a function of distance from cluster centres, for the two samples at $z\sim 0.2$ (left panel) and $z\sim 0.3$ (right panel), again split into three stellar mass bins. Symbols represent cluster values, while lines represent field values for the corresponding stellar mass.  The blue fraction decreases with stellar mass.  At fixed stellar mass,  the change of the blue fraction happens within $\sim 3$Mpc from the cluster centres, with no significant further changes beyond $\sim 3$ Mpc. See text for a detailed discussion.\label{fred}}
\end{figure*}

\begin{figure*}  
\includegraphics[width=0.48\textwidth]{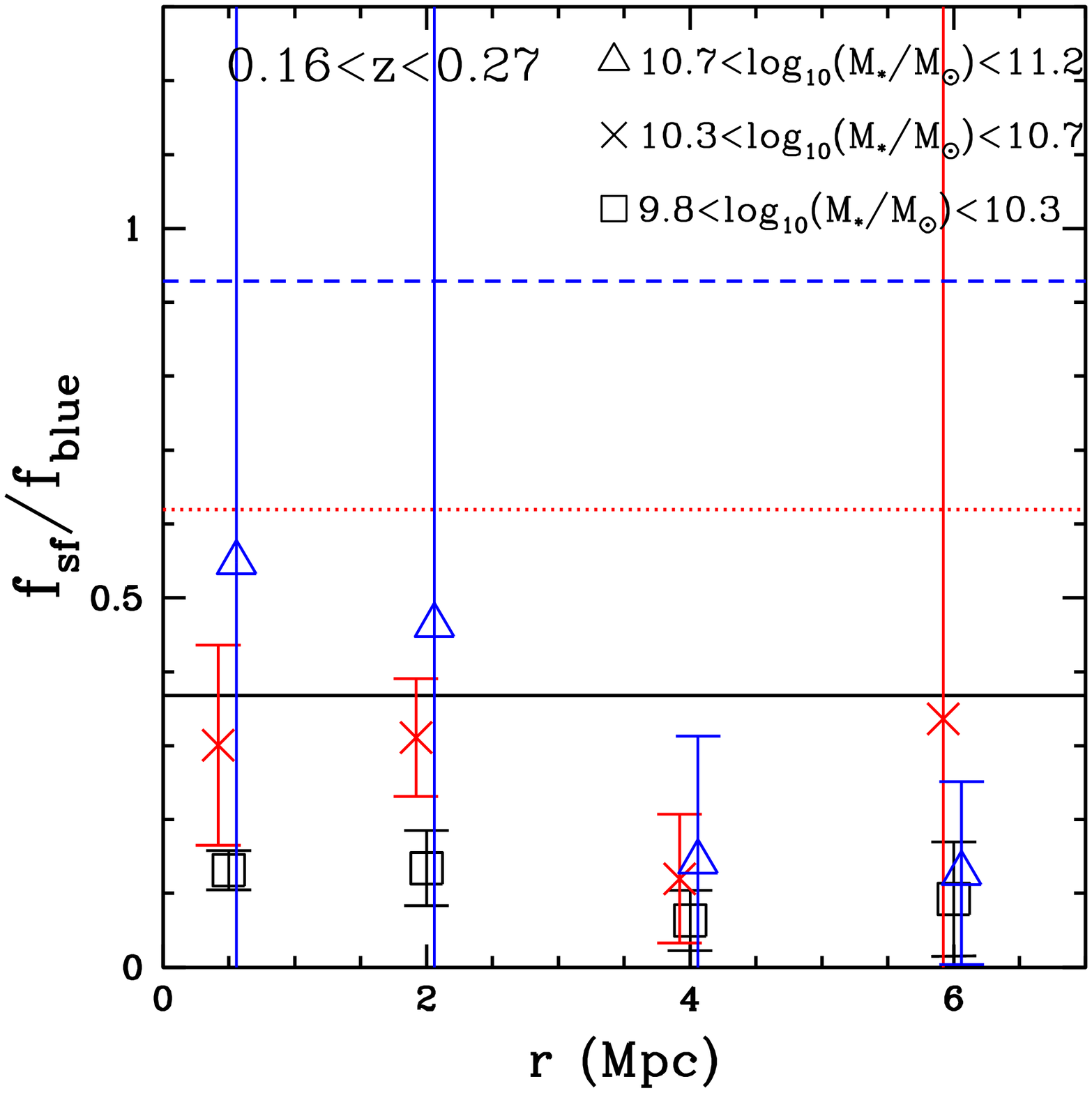}\includegraphics[width=0.48\textwidth]{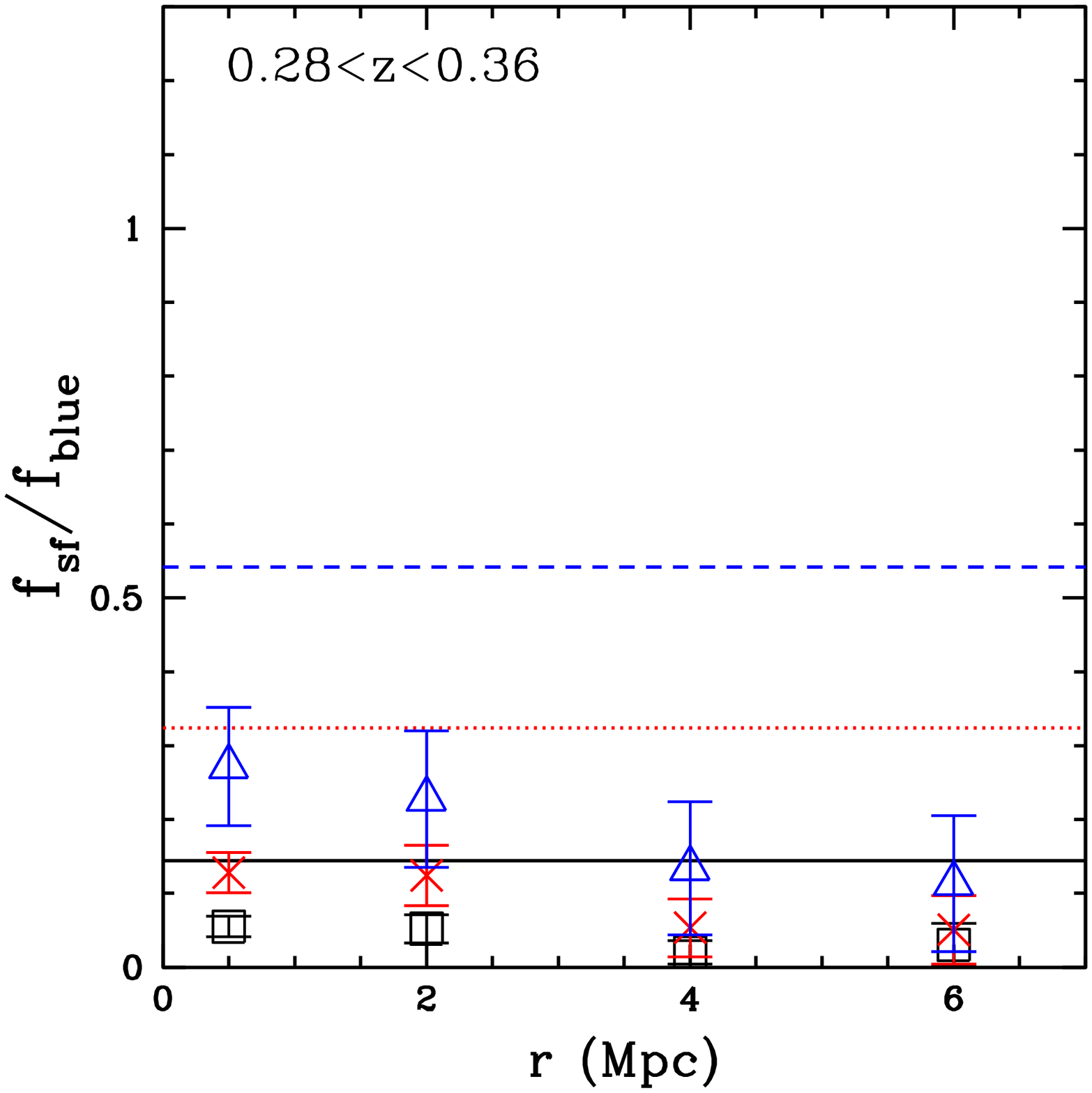}
\caption[]{Fraction of blue galaxies that have SFR above our UV detection limit. The two panels are for two redshifts. Different symbols represent cluster galaxies with different stellar masses, and the lines represent field galaxies at corresponding stellar masses. }\label{fsfblue}
\end{figure*}

To estimate the fraction of galaxies that are on the red-sequence, we refine the model slope of the red-sequence from our data directly. We stack clusters in each redshift bin to a central redshift (passively evolved based an old SSP \citealt{galexev} model), and fit the red-sequence and the scatter, $\sigma$, around the fitted colour-magnitude relation.

We consider galaxies that are redder than 3$\sigma$ below the fitted red-sequence as red galaxies\footnote{We have also confirmed that if defining red galaxies by mirroring the redder half of the fitted red-sequence \citep{lu09} our conclusions do not change.}. The fraction of blue galaxies as a function of distance from cluster centres is plotted in Figure \ref{fred}, for the two samples at $z\sim 0.2$ (left panel) and $z\sim 0.3$ (right panel), again split into three stellar mass bins (black squares, red crosses, and blue triangles). At both redshifts, the blue fraction decreases with stellar mass, at all radii. At fixed stellar mass, it is quite clear that at both redshifts, the change of the blue fraction happens within $\sim 3$ Mpc from the cluster centres, with no significant further changes beyond $\sim 3$ Mpc. Furthermore, the difference between the blue fraction in the outer and inner most regions is smaller for the most massive galaxies compared to those of lower masses.

It is interesting also to compare this with the field values. We calculate the blue fraction for the phot-z field sample in a similar fashion as for the cluster sample, i.e. red galaxies are defined as those  that are redder than 3$\sigma$ below the fitted field red-sequence. The resulting field blue fractions are plotted in Figure  \ref{fred} as the black solid, red dotted, and blue dashed lines for the three stellar masses respectively.  Not surprisingly, in the central regions of the clusters ($r<1$ Mpc) the blue fraction is lower than that in the field.  The more interesting thing is what happens in the outer regions. Beyond $r\sim 3$ Mpc, the cluster blue fraction is generally lower than that in the field, perhaps surprising given the low density contrast at these radii. Only the most massive galaxies, at $z\sim 0.2$, reach blue fractions comparable to the field at these radii.  Keep in mind that due to the nature of the background subtraction, what we detect is the ``excess'' over the field population (at $r\sim 7$ Mpc, the excess is on the $\sim$10 per cent level with respect to the density in the field).  We discuss the implication of these results in combination with the other results presented above in the next Section.

\section{Discussion}\label{gdis}
Our results show that the fraction of blue, actively star-forming 
galaxies is significantly lower in dense environments, even far from
cluster cores.  However, no significant difference in the distribution of SFR for the active
population (with SFR above our SFR$_{\text{UV}}$ detection limits) is detected. 
  Both the average SFR and the shape of
the distribution are the same in the field, the dense cluster cores,
and overdense regions in the distant outskirts.  This is in good
agreement with a growing body of independent results.
Studying the local universe, several authors have shown that the distribution of colour
and/or SFR for star--forming galaxies in
the SDSS is independent of environment \citep{Balogh2004,
  peng10,wetzel11}.  Recently, \citet{biv11} came to the same
conclusion using 24$\mu$m--derived SFRs around a $z=0.23$ supercluster;
while the fraction of star-forming galaxy depends sensitively on
environment, in a complex way, the correlation between SFR and stellar mass 
remains unchanged.  At even higher redshifts of $0.3<z<0.5$,
\citet{mcgalex10} also found the mean SFR of star--forming galaxies in
groups to be independent of environment.  Notably, the results of
\citet{peng10} and \cite{mcgalex10} extend to SFR lower than we probe
in this paper, and yet they still find no population of low-SFR
galaxies in clusters.

This is in apparent contrast with the work by \cite{vul2010}, who
measured SFR from [OII] and 24$\mu$m data in  the ESO Distant Cluster
Survey, reaching a SFR depth comparable to ours. They found that the average SFR of star-forming galaxies in these $z\sim 0.5$ clusters is a factor of 1.5 lower than that in the field. 
One possibile explanation for the discrepancy with our data is that if the low SFR population in clusters are
sufficiently dusty \citep[e.g.][]{wolf09}, they might not make it into
our NUV--selected sample. Galaxies at the limiting apparent NUV magnitude in our
sample and with extinction greater than $\sim 1-2$ mag would
not have made it into our final SFR limited sample.

We can further gain insight into the
population with SFR below our UV detection limit (possibly including
the dusty ones) by comparing the fraction 
of star forming galaxies among the optically blue population. In Figure
\ref{fsfblue} we plot the fraction of blue galaxies that have SFR above
our UV detection limit. The two panels are for two redshifts. Different
symbols represent cluster galaxies with different stellar masses, and
the lines represent field galaxies at corresponding stellar masses. As
expected, not all blue galaxies are star forming (at a rate higher than
our relatively high UV detection limit). However, what is especially remarkable is
the difference between the field and cluster galaxies. Compared to the
blue galaxies in the field, a lower fraction of the blue galaxies in
the clusters is forming stars at a rate above our UV detection limit. This
indicates a larger population of blue galaxies with low SFR (lower than
our UV detection limit) in clusters than in the field. Although, it is possible that, as mentioned above, these blue galaxies  in clusters are dustier than their field counterparts \citep{wolf09,haines11}, and thus would still be optically blue but suppressed in the UV. If indeed their SFR is reduced and yet they are still blue, it would require a quenching mechanism that does not completely shut
down the SFR on short timescales. Further, this mechanism must affect
galaxies in a way that is independent of proximity to the cluster core, as this fraction is roughly constant with radius within the uncertainties.

\begin{figure*} 
 \includegraphics[width=0.9\textwidth]{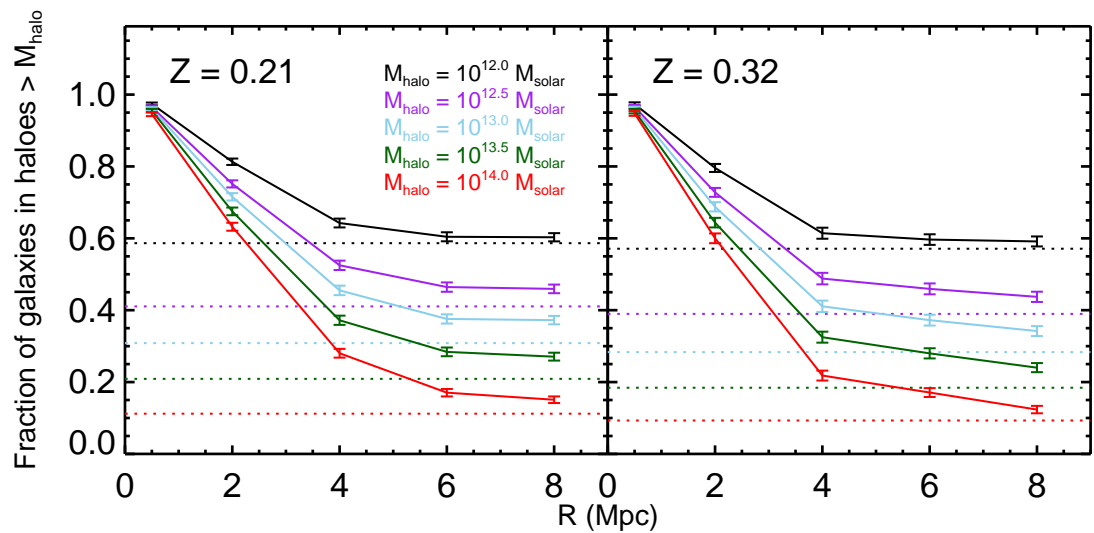}   \caption[]{Fraction of galaxies ($9.8<\log(\text{M}_*/\som)<10.3$) as a function of cluster-centric radius which reside in halos above a given mass in the semi-analytic models of \cite{Font2008}. Solid curves, and dashed lines indicate cluster and field values respectively. \label{f_sim}}
\end{figure*}
It is also notable that, at the outskirts ($\sim 7$Mpc), despite the
low average density contrast with the field, cluster galaxies have a
lower blue fraction than that in the field (except, perhaps, for the most massive galaxies at z$\sim 0.2$).  This indicates that those galaxies that just arrived at the cluster environment already have their star formation quenched, supporting a ``pre-processing" scenario suggested by several studies \citep[e.g.][]{zm98,B00,BM10,mcgeeacc}. The recent work by \cite{wetzel11} also found enhanced fraction of quenched galaxies out to $\sim 10 r_{200}$ in SDSS groups/clusters  and  attributed it to satellite galaxies that have possibly been pre-processed in groups.  To gain more insight into this issue, we use the semi-analytic models of \cite{Font2008} to construct the halo mass function of galaxies at different cluster-centric radii and compare it with that of the field galaxies. The \cite{Font2008} model is a recent version of the GALFORM model \citep{Cole2000,bowermodel} which has been modified to include a more realistic treatment of environmental effects. Although this method still has some problems in reproducing the distribution of galaxy colours \citep[e.g.][]{Balogh2009}, the stellar mass functions agree  reasonably well with observations to $z\sim5$. To mimic our observations of the large scale environment around clusters, we construct a stacked model cluster by selecting all galaxies within $\pm$ 4000 km/s of the central galaxy of clusters ($M_h > 10^{14.2}\ \som$) in the Font model. The advantage of using the models is that the host halo mass of each galaxy is known.  In Figure \ref{f_sim}, we show the resulting fraction of galaxies with stellar masses of $9.8<\log(\text{M}_*/\som)<10.3$ as a function of cluster-centric radius which reside in halos above a given halo mass. The lower and higher redshift bins are presented in the left and right panels respectively. Solid curves present cluster values, and the dashed lines indicate field values (see legend for details). It shows that at around $\sim 6$ Mpc, the outermost region we probed with our data here, the fraction of galaxies residing in halos above $10^{12.5}\ \som$ is slightly higher than that in the field. This could be part of the explanation of our results: the higher red fraction in the outskirts is due to the fact that more galaxies are in halos above group scale halos in the outskirts than in the field, provided that the quenching already started in those halos. However, it is not clear whether those galaxies residing in groups in the outskirts are further processed in the proximity of the large-scale cluster environment. Furthermore, because of the background subtraction, what we observe is the properties of the cluster population in ``excess'' of the field. Thus, it is hard from our data to  determine the exact nature of the excess we detected at large radii.  A comparison between groups in the outskirts and isolated groups in the field will provide more insight. Therefore, we defer a closer examination of groups in the outskirt regions to a future paper, where we use spectroscopic data to study two contrasting clusters.

\section{Conclusions}\label{con}

In this work, we examined the star formation properties  of a large sample of $\sim 100$ galaxy clusters at $0.16<z<0.36$, from their  cores out to $\sim 7$ Mpc, using the CFHTLS optical data and GALEX UV data. Our main findings are summarized below.

(i) We found that the average SSFR  and the distribution of SSFR (of galaxies with SFR$>0.7\ \som$/yr at $z\sim 0.2$ and  SFR$>1.2\ \som$/yr at $z\sim 0.3$) show no strong dependence on the distance from the cluster centre within the error bars, and are similar to that in the field as well. 

(ii) The fraction of star-forming galaxies  (with SFR$>0.7\ \som$/yr at $z\sim 0.2$ and  SFR$>1.2\ \som$/yr at $z\sim 0.3$) is much lower in clusters than in the field. For cluster galaxies, this star-forming fraction is  radius and stellar mass independent within the uncertainties, partially due to our relatively high SFR$_{\text{UV}}$ limits.

(iii) Among the optically blue population in clusters, a lower fraction  is forming stars at a rate higher than our SFR$_{\text{UV}}$ limits compared to the blue population in the field. This difference is larger for high mass galaxies, and is roughly independent of radius.

(iv) The fraction of galaxies with blue colours is constant from $\sim 3$ Mpc out to $\sim 7$ Mpc; however, within 3 Mpc, there is an abrupt decrease in this fraction towards the cluster core. This is present at both redshifts, at all stellar masses examined here, but more so for the least massive galaxies ($9.8<\text{log}_{10}(\text{M}_*/\som)<10.3$).

(v) Despite the low average density contrast with the field in the outermost region ($r\sim 7$ Mpc), the blue fraction  is lower than that in the field; with the exception of the most massive galaxies at $z\sim 0.2$.

Our results imply that the excess population over the field in the
outskirts of clusters is pre-processed, and at all radii throughout the
clusters there is a population of blue galaxies that have their SFR
reduced to below our UV detection limit but not to zero. This requires a
mechanism that does not shut off the star formation completely, and
works in a way that is independent of radius. Limited by our SFR$_{\text{UV}}$ 
detection limit, we cannot probe the distribution of the SFR of that
partially-quenched population. With deeper data to detect these
galaxies and measure the change in the shape of the SFR distribution,
it may be possible to put stronger constraints on the timescale on
which the quenching mechanism operates.

\section*{Acknowledgment}
 The authors thank the other members of the CLUE collaboration (Marcin Sawicki, Mike Hudson and Brian McNamara) for their contributions. We also thank the referee for suggestions that improved the presentation of the paper. This work was supported by an Early Researcher Award from the province of Ontario and an NSERC Discovery Grant to MLB, and was made possible by the facilities of the Shared Hierarchical Academic Research Computing Network (SHARCNET:www.sharcnet.ca). SLM acknowledges support from a NSERC Postdoctoral Fellowship.

Based on observations obtained with MegaPrime/MegaCam, a joint project of CFHT and CEA/DAPNIA, at the  Canada-France-Hawaii Telescope (CFHT) which is operated by the National Research Council (NRC) of Canada, the Institut National des Science de l'Univers of the Centre National de la Recherche Scientifique (CNRS) of France, and the University of Hawaii. This work is based in part on data products produced at TERAPIX and the Canadian Astronomy Data Centre as part of the Canada-France-Hawaii Telescope Legacy Survey, a collaborative project of NRC and
CNRS.


\appendix

\section{Cross-matching Catalogues}

\begin{figure} 
\begin{center}
\includegraphics[width=0.5\textwidth]{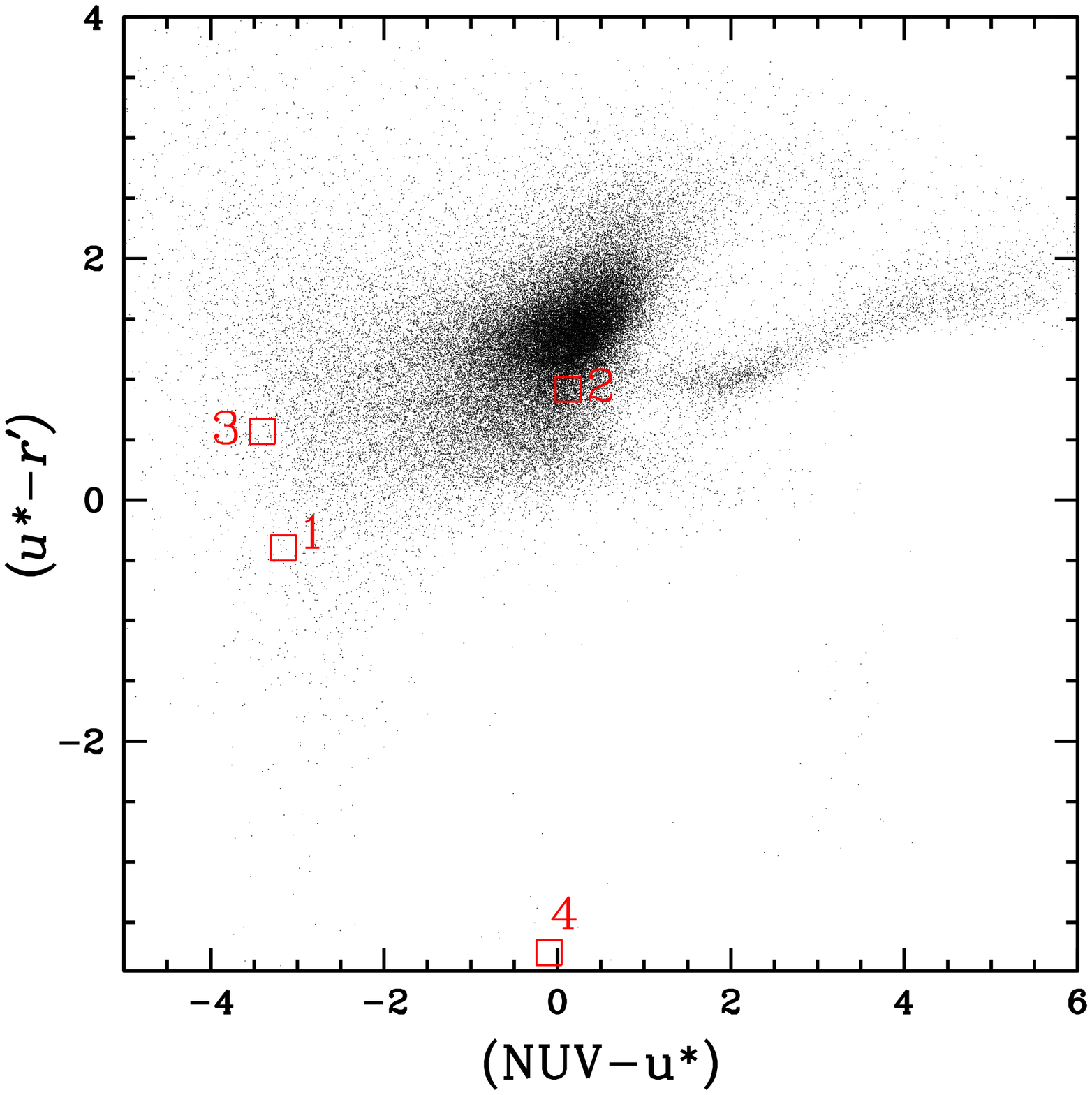} \caption[$u^*-r'$ vs $NUV-u*$ for cross-matching optical and NUV sources]{One example of how we use colour information to cross-match NUV sources with optical sources. Small dots are galaxies at  $22.5<NUV<23.0$ and  have only one possible optical counterpart. There is an obvious concentration of dots in a specific region in this colour-colour space, indicating the colours the real matches are mostly likely to have. The four squares are four possible optical matches within 4 arcsec from a NUV source with  NUV=22.5, numbered in the order of distance from the NUV source, with 1 being the closest one. It shows that the one that is the closest match in this case is less likely to be the real match than the second closest match, because it is located in a less dense region.
\label{c-c-mat}}
\end{center}
\end{figure}

As discussed in Section \ref{cross}, in cases where there are one or more candidate optical matches within 1 arcsec of the closest match to a NUV source, we use colours to help identify the most likely optical counterpart. We do this using galaxies at fixed NUV magnitude. For a galaxy at a certain magnitude, the probability of it having a certain colour  is not random. In Figure \ref{c-c-mat}, the $(u^*-r')$ colours of the objects that only have one possible optical counterpart (small dots) in the magnitude range  $22.5<NUV<23.0$ are plotted against their  $(NUV-u^*)$ colours. As we can see, the density of the dots in this colour-colour plane is not uniform; instead, most galaxies have colours of $(u^*-r')\sim 1$ and  $(NUV-u^*)\sim 0$, and we interpret this as the most probable colour for a galaxy of this magnitude. Therefore, when there are multiple possible optical matches for a NUV source, we take the one that resides in the region with the highest density as the real match.  One example is shown in Figure \ref{c-c-mat}. The four squares are four possible optical matches to a NUV source with  NUV=22.5 mag, numbered in the order of distance from the NUV source, with 1 being the closest. The one that is the closest match is residing in a region where the density of the points is less than that of the second closest match, and thus in this case we take the spatially second closest match as the real match. In about 15 percent of the cases, the spatially closest match is different from the colour-based match.

\end{document}